\documentclass[10pt]{article}
\usepackage{amsmath,amsthm,latexsym,amssymb,amsfonts,epsfig,psfrag}
\usepackage{color}
\usepackage{hyperref}
\usepackage{tikz}
\usetikzlibrary{patterns}

\makeatletter \@addtoreset{equation}{section} \makeatother

\def\be{\begin{equation}}
\def\ee{\end{equation}}
\def\ba{\begin{eqnarray}}
\def\ea{\end{eqnarray}}

\newcommand\nn{\nonumber}
\newcommand\q{\quad}
\def\Nl{{\mathchoice
{\setbox0=\hbox{$\displaystyle\rm N$}\hbox{\hbox to0pt
{\kern0.4\wd0\vrule height0.9\ht0\hss}\box0}}
{\setbox0=\hbox{$\textstyle\rm N$}\hbox{\hbox to0pt
{\kern0.4\wd0\vrule height0.9\ht0\hss}\box0}}
{\setbox0=\hbox{$\scriptstyle\rm N$}\hbox{\hbox to0pt
{\kern0.4\wd0\vrule height0.9\ht0\hss}\box0}}
{\setbox0=\hbox{$\scriptscriptstyle\rm N$}\hbox{\hbox to0pt
{\kern0.4\wd0\vrule height0.9\ht0\hss}\box0}}}}
\def\Zl{{\mathchoice
{\setbox0=\hbox{$\displaystyle\rm Z$}\hbox{\hbox to0pt
{\kern0.4\wd0\vrule height0.9\ht0\hss}\box0}}
{\setbox0=\hbox{$\textstyle\rm Z$}\hbox{\hbox to0pt
{\kern0.4\wd0\vrule height0.9\ht0\hss}\box0}}
{\setbox0=\hbox{$\scriptstyle\rm Z$}\hbox{\hbox to0pt
{\kern0.4\wd0\vrule height0.9\ht0\hss}\box0}}
{\setbox0=\hbox{$\scriptscriptstyle\rm Z$}\hbox{\hbox to0pt
{\kern0.4\wd0\vrule height0.9\ht0\hss}\box0}}}}
\def\Ql{{\mathchoice
{\setbox0=\hbox{$\displaystyle\rm Q$}\hbox{\hbox to0pt
{\kern0.4\wd0\vrule height0.9\ht0\hss}\box0}}
{\setbox0=\hbox{$\textstyle\rm Q$}\hbox{\hbox to0pt
{\kern0.4\wd0\vrule height0.9\ht0\hss}\box0}}
{\setbox0=\hbox{$\scriptstyle\rm Q$}\hbox{\hbox to0pt
{\kern0.4\wd0\vrule height0.9\ht0\hss}\box0}}
{\setbox0=\hbox{$\scriptscriptstyle\rm Q$}\hbox{\hbox to0pt
{\kern0.4\wd0\vrule height0.9\ht0\hss}\box0}}}}
\def\Rl{{\mathchoice
{\setbox0=\hbox{$\displaystyle\rm R$}\hbox{\hbox to0pt
{\kern0.4\wd0\vrule height0.9\ht0\hss}\box0}}
{\setbox0=\hbox{$\textstyle\rm R$}\hbox{\hbox to0pt
{\kern0.4\wd0\vrule height0.9\ht0\hss}\box0}}
{\setbox0=\hbox{$\scriptstyle\rm R$}\hbox{\hbox to0pt
{\kern0.4\wd0\vrule height0.9\ht0\hss}\box0}}
{\setbox0=\hbox{$\scriptscriptstyle\rm R$}\hbox{\hbox to0pt
{\kern0.4\wd0\vrule height0.9\ht0\hss}\box0}}}}
\def\Cl{{\mathchoice
{\setbox0=\hbox{$\displaystyle\rm C$}\hbox{\hbox to0pt
{\kern0.4\wd0\vrule height0.9\ht0\hss}\box0}}
{\setbox0=\hbox{$\textstyle\rm C$}\hbox{\hbox to0pt
{\kern0.4\wd0\vrule height0.9\ht0\hss}\box0}}
{\setbox0=\hbox{$\scriptstyle\rm C$}\hbox{\hbox to0pt
{\kern0.4\wd0\vrule height0.9\ht0\hss}\box0}}
{\setbox0=\hbox{$\scriptscriptstyle\rm C$}\hbox{\hbox to0pt
{\kern0.4\wd0\vrule height0.9\ht0\hss}\box0}}}}
\def\Hl{{\mathchoice
{\setbox0=\hbox{$\displaystyle\rm H$}\hbox{\hbox to0pt
{\kern0.4\wd0\vrule height0.9\ht0\hss}\box0}}
{\setbox0=\hbox{$\textstyle\rm H$}\hbox{\hbox to0pt
{\kern0.4\wd0\vrule height0.9\ht0\hss}\box0}}
{\setbox0=\hbox{$\scriptstyle\rm H$}\hbox{\hbox to0pt
{\kern0.4\wd0\vrule height0.9\ht0\hss}\box0}}
{\setbox0=\hbox{$\scriptscriptstyle\rm H$}\hbox{\hbox to0pt
{\kern0.4\wd0\vrule height0.9\ht0\hss}\box0}}}}
\def\Ol{{\mathchoice
{\setbox0=\hbox{$\displaystyle\rm O$}\hbox{\hbox to0pt
{\kern0.4\wd0\vrule height0.9\ht0\hss}\box0}}
{\setbox0=\hbox{$\textstyle\rm O$}\hbox{\hbox to0pt
{\kern0.4\wd0\vrule height0.9\ht0\hss}\box0}}
{\setbox0=\hbox{$\scriptstyle\rm O$}\hbox{\hbox to0pt
{\kern0.4\wd0\vrule height0.9\ht0\hss}\box0}}
{\setbox0=\hbox{$\scriptscriptstyle\rm O$}\hbox{\hbox to0pt
{\kern0.4\wd0\vrule height0.9\ht0\hss}\box0}}}}
\newcommand{\bd}{\mathbf d}

\title{Time evolution as refining, coarse graining and entangling}
\author{Bianca Dittrich,  Sebastian Steinhaus\\
\small  Perimeter Institute for Theoretical Physics,\\
 \small 31 Caroline Street North, Waterloo, Ontario, Canada N2L 2Y5
}

\date{}

\begin{document}

\maketitle

\begin{abstract}
We argue that refining, coarse graining and entangling operators can be obtained from time evolution operators. This applies in particular to geometric theories, such as spin foams. We point out that this provides a construction principle for the physical vacuum in quantum gravity theories and more generally allows to construct a (cylindrically) consistent continuum limit of the theory.

\end{abstract}

%\tableofcontents

\section{Introduction}

Renormalization and coarse graining have become  powerful tools  to connect   microscopic and macroscopic regimes of a given theory.
In particular many approaches to quantum gravity postulate or aim to derive macroscopic space time as arising from the collective dynamics of basic building blocks \cite{lollreview,eckert,gfthydro,livine}. To validate such a picture one has to show that the many body dynamics of such systems gives indeed a smooth space time if sufficiently coarse grained.

To this end coarse graining techniques \cite{wilson} need to be employed. The question of how to coarse grain or block fine degrees of freedom into coarser ones is essential for determining good truncations for the coarse graining schemes. Coarse graining maps are dual to refining maps, in fact  tensor network renormalization schemes \cite{levin,guwen} put the emphasis rather on refining maps, that then also determine the properties of the truncation in these schemes \cite{bd12b}.

 %The tensor network coarse graining \cite{levin,guwen,vidal}, leads to a connection between the covariant and canonical picture, see for instance \cite{bd12b}.  It provides coarse graining and refining maps (and (dis-) entangling maps in \cite{vidal}) for the canonical theory. The refining maps lead to embedding maps which can be used to construct the continuum limit of the theory via an inductive limit. This is a standard construction in loop quantum gravity, which currently is based on a kinematical embedding map,  describing the (kinematical) Ashtekar--Lewandoswki vacuum \cite{al}. This way of defining a continuum limit can then be compared to constructing the macroscopic dynamics via tensor network coarse graining, as is done in \cite{bd12b}. As is argued there the key difference is in the choice of embedding or refining maps. 

%\cite{bd12b} suggested to change the kinematical embedding maps, employed in the kinematics of loop quantum gravity, to embeddings which are determined by the dynamics of the theory and are obtained from the refining maps appearing in the tensor network coarse graining algorithms.

 In this document we point out that time evolution maps, which appear in simplicial discretizations \cite{hoehn2,hoehn3},  can also be interpreted as refining and coarse graining maps.  As we will argue here this applies in particular to gravitational dynamics, e.g. spin foams \cite{ponzanoregge,alexreview,cbook,eprl}.  
 
 %We will show that such refinement maps allow the consistent definition of the dynamics of the continuum theory via a so--called inductive limit construction. 

%Additionally, time evolution maps  between equal size\footnote{This assumes finite dimensional Hilbert spaces. Even for infinite dimensional Hilbert spaces we can make `size' more precise: In a discrete dynamics, Hilbert spaces carrying the degrees of freedom, are associated to sites, edges or other geometrical objects. The Hilbert space describing the  states at a given time is then (typically) given as a tensor product of these basic Hilbert spaces.  `Size' then refers to the complexity of the underlying discretization. This notion can be made concrete by introducing a partial order on the set of discretizations. In simple cases `size' just refers to the number of sites.} Hilbert spaces can be seen as entangling maps, which is a quite common notion. Again, we will argue that gravitational systems are also special in this regard. 

One reason why the appearance of time evolution as coarse graining or refining maps applies in particular to gravitational or other diffeomorphism invariant systems is the following: As argued in \cite{rocek,louapre,bd08,bahrdittrich09a,steinhaus11} diffeomorphism symmetry in discrete systems translates to a symmetry, which can be interpreted as moving vertices in the discrete space time described by the dynamical variables of the theory. These vertex translations can also be understood as time evolution. Now, vertices can be even moved on top of each other, which gives a coarse graining of the underlying state. Alternatively vertices can split into two and in this way give a refinement. Indeed this argument was used in \cite{steinhaus11} to show that diffeomorphism symmetry implies discretization independence.  

%We will present in section \ref{scalar} an example where time evolution is also given by vertex (or edge) translations forward in time. In this case a time evolution move is necessarily connected to a splitting of two vertices and hence a refinement of the underlying discretization.

More generally diffeomorphism invariant systems are totally constrained, i.e. the Hamiltonian is given by a combination of constraints. %Time evolution in the discrete system can be described by a transfer matrix or time evolution operator -- which is nothing else than a partition function with two boundaries representing the two time slices.
 In the case of a totally constrained system the time evolution operator should be a projection operator \cite{proj,alex3d}, projecting onto so--called physical states. Thus physical  states should not evolve.\footnote{Introducing relational observables, a notion of relational time evolution can be reconstructed \cite{partial}. In this paper we mean with time evolution always evolution with respect to (unphysical) coordinate time, which just acts as gauge transformation and hence acts as an identity on physical states.}

For discrete time evolutions that change the number of degrees of freedom, this leads to the puzzle of how to identify states from Hilbert spaces of `different size'.\footnote{This assumes finite dimensional Hilbert spaces. Even for infinite dimensional Hilbert spaces we can make `size' more precise: In a discrete dynamics, Hilbert spaces carrying the degrees of freedom, are associated to sites, edges or other geometrical objects. The Hilbert space describing the  states at a given time is then (typically) given as a tensor product of these basic Hilbert spaces.  `Size' then refers to the complexity of the underlying discretization, that is the number of sites, edges etc.}
  We will argue that such states describe indeed the same physical state, however expressed on two different discretizations. The equivalence relation is provided by the refining time evolution operator. We will explain how this notion can be formalized into the construction of an inductive limit Hilbert space. Such an inductive limit construction is also used for the (kinematical) Hilbert space of loop quantum gravity \cite{al,thomasbook}.

The inductive limit Hilbert spaces, which are defined via an equivalence relation between states from Hilbert spaces based on different discretizations, require however (so called cylindrical)  consistency conditions: physical observables should not depend on which representative they have been determined on. Indeed we will connect these consistency conditions with a notion of path independence for (refining) time evolution. This  relates then to the requirement of diffeomorphism invariance.

Discrete (non--topological) theories typically break the diffeomorphism symmetry \cite{bahrdittrich09a}. The hope however is that diffeomorphism symmetry can be recovered in the continuum limit. We will explain how to formulate the continuum limit of the dynamics of a given quantum gravity theory and how such a continuum limit can be constructed by an iterative coarse graining procedure akin to tensor network renormalization schemes. 

%This is known as cylindrical consistency, which indeed restricts the possible observables of the theory. If such a notion of cylindrical consistency is based on the dynamics \cite{bd12}, one would expect that cylindrical consistency conditions for observables are related to the requirement of Dirac (or diffeomorphism invariant) observables.  Indeed we will argue that this is the case due to the fact that time evolution is deeply entwined with refining maps. 

Topological theories can be often discretized without breaking diffeomorphism symmetry. In this case refining time evolution maps indeed satisfy the consistency conditions. We point out that this provides a construction principle for inductive limit Hilbert spaces, that can for instance be applied to find new quantum representations for loop quantum gravity \cite{geiller}.

The idea that time evolution can be interpreted  as coarse graining, refining or entangling occurs in many approaches. Tensor network coarse graining algorithms can be easily seen as time evolution in radial direction (in an Euclidean space time), which itself leads to holographic renormalization \cite{holographic}.  Entanglement  renormalization \cite{vidal}, which is also based on tensor network techniques,  can be interpreted in a space time picture, again involving holographic renormalization, see for instance  \cite{swingle}.  Here the tensor network and the entanglement it encodes are interpreted as a (background) AdS space time. Although such geometrical interpretations appear very naturally, the interpretation of the underlying geometry as a background geometry might not apply straightforwardly to gravity. The reason is that the dynamical variables include the geometric degrees of freedom. Hence the geometry is encoded in the boundary state itself, and has to be extracted from it. 

A main point of this paper is to bring together  coarse graining tools developed in condensed matter with methods developed in loop quantum gravity and to point out the many peculiarities that arise if one considers totally constrained systems such as general relativity. This leads to our proposal of how to construct the continuum limit of a given quantum gravity theory, together with a notion of a physical vacuum state. Furthermore we point out a general construction principle for inductive limit Hilbert spaces based on time evolution maps of topological theories.

\subsection{Overview}

In this paper we will employ a generalized meaning of time evolution, which will be explained in sections \ref{class} and \ref{timeevol}. The first generalization applies in particular to discretized field theories, where we allow for a time evolution, which changes the number of variables, that is phase space or Hilbert space dimension, from one time step to the next. The second issue we will discuss, is the meaning of time evolution in a totally constrained system, such as general relativity.

Usually one considers a discretization that does not change in time, and thus the number of degrees of freedom stays also constant. However, for theories involving a curved background, or gravity as a dynamical theory, one often  uses an irregular lattice, where the discretization and the number of variables do change in time. %Consider for instance the evolution of a discrete scalar field in an expanding universe \cite{foster}. 

In section \ref{class} we will discuss time evolution in simplicial discretizations, where in general the number of degrees of freedom change. Such simplicial discretizations are in particular used for (the quantization of) gravity, for instance in Regge calculus \cite{regge} or spin foams \cite{cbook}. 

The quantization of the Hamiltonian constraint in loop quantum gravity \cite{thiemann} also involves a change of the underlying discretization (in the form of a graph). The interpretation of this graph changing Hamiltonian is an open issue. In this work we will suggest an interpretation for a graph or discretization changing time evolution. On the other hand this interpretation will help to actually design reasonable discrete dynamics involving a change of phase or Hilbert space.

In  sections \ref{class}  and \ref{quant} we will also explain how to formulate such a dynamics with varying number of degrees of freedom in the classical and quantum realm respectively and propose that such a dynamics can be interpreted to refine  or coarse grain a given state. This is underlined with a number of examples in section \ref{class}. 

This interpretation is strengthened if we consider totally constrained systems, such as general relativity or topological field theories. In a totally constrained system the Hamiltonian is given as a combination of constraints $C_i$, that generate gauge transformations. Thus time evolution is equivalent to a gauge transformation, realizing the fact that in such systems the choice of time coordinate is arbitrary. 

The classical evolution of such systems does not change the states, as these are defined as gauge equivalence classes. The quantum evolution in the form of a path integral
\ba\label{pathintegral}
 \int_{X_{ini},X_{fin}} {\cal D} X \, \exp\left( \frac{i}{\hbar} S(X) \right)
\ea
is supposed to act as a  projector onto physical states $\psi_{phys}(X)$ annihilated by the quantized constraints $\hat C_i \psi_{phys}=0$ \cite{proj}.  Thus evolution with respect to (coordinate) time is `frozen'. 

Consider a discretization of a totally constrained system and allow for the number of degrees of freedom to change during time evolution. Here we will understand time evolution in the sense of (\ref{pathintegral}), that is we consider a discretized path integral. How should we interpret this time evolution, which is supposed to be `frozen', in the case that the number of variables involved (including physical and gauge degrees of freedom) does change?

We will propose in section \ref{quant} that in this case time evolution is equivalent to a refining or coarse graining of a state. (In case the initial state is not physical, unphysical degrees of freedom might be also projected out.) We will connect the case of refining time evolution to the construction of a continuum Hilbert space via an inductive limit, explained in section \ref{emb}, as is used in loop quantum gravity \cite{al}. Such a construction provides a precise sense in which states from Hilbert spaces of `different size' can be equivalent. Note that this inductive limit construction for the continuum Hilbert space has  so far been used only for the kinematical Hilbert space in loop quantum gravity. We propose here a construction which involves the dynamics.  Thus the dynamics defines which states are equivalent, as one would expect for the physical Hilbert space, i.e. the space of states, satisfying the constraints.

Considering a time evolution where the number of degrees of freedom change, we can go to the extreme, and start from an `empty' discretization, supporting no variables at all. This will be discussed in section \ref{timeevol}. A state resulting from such a refining time evolution with such initial conditions defines the (Hartle--Hawking) no--boundary state. The different stages of evolution just represent this  state on different discretizations, which is consistent with the construction of a Hilbert space via an inductive limit.  We will propose that refining a given  state via time evolution, means to put the additional degrees of freedom into a state, that resembles this Hartle--Hawking  state in some localized form. It is thus natural to see the Hartle--Hawking state as the vacuum state of the system. (Note that in constrained systems the definition of vacuum via minimal energy is usually not available - all states satisfy the Hamiltonian constraints and have therefore zero energy, at least in systems without a boundary.)

Often discretizations provide the only method to make sense of the formal continuum path integral. However for (non--topological) systems the continuum diffeomorphism symmetry is typically broken by the discretization \cite{bahrdittrich09a}. But without a realization of diffeomorphism invariance in the path integral (\ref{pathintegral}), it cannot act as a projector onto the physical states. 
To deal with this issue one attempts  to restore diffeomorphism invariance via refining the building blocks and finding effective amplitudes for the coarser building blocks by integrating out the finer degrees of freedom \cite{nielsen,bahrdittrich09b}. This we usually refer to as coarse graining flow (although the initial step is a refining). We will explain how this defines a continuum limit of (\ref{pathintegral}), which can be expressed as a cylindrically consistent amplitude map on an inductive limit Hilbert space in section \ref{howto}. For such an  inductive limit Hilbert space one needs to again define refinement maps, which we propose to be given by (effective) time evolution maps. This holds in particular if one wants to express the physical Hilbert space as an inductive limit.

Thus what has been said above about equivalence of time evolution and refining and coarse graining will hold in general only in some approximate sense. In fact, one can now attempt to construct discretizations for which this holds to a good approximation. This will also provide the means to define the continuum limit via a coarse graining flow. In this continuum limit one then expects this equivalence to hold exactly.

Section \ref{tensor} will explain that tensor network renormalization schemes provide a means to construct cylindrically consistent amplitude maps and an inductive limit physical Hilbert space. On the other hand the insight that time evolution maps provide refining maps might help to develop new tensor network renormalization schemes.

The breaking of diffeomorphism symmetry by discretizations can often be avoided in topological theories, such as three--dimensional gravity. Here the relation between time evolution and refining or coarse graining can be made exact.  We will therefore illustrate our claims with examples from topological field theories in section \ref{top}. In particular the (refining) time evolution maps can be taken as refinement maps for the construction of an inductive limit Hilbert space. Note that the applicability of this idea is not exclusive to topological field theories: one can also use the time evolution maps of topological field theories to define inductive limit Hilbert spaces for other theories.  Based on this idea a new representation for loop quantum gravity has been recently defined in \cite{geiller}, based on the time evolution maps for $BF$--theory.  We will explain in section \ref{subtop} that this construction can be generalized to other (discretized) topological field theories. It provides a method to find Hilbert space representations for non--topological theories based on vacua provided by the topological theories.

Section \ref{geom} will comment more on the peculiarities in gravitational theories, where the geometric scale is part of the dynamical variables. It will provide a geometric interpretation of the refining time evolution maps and make clear that these should indeed be rather seen as refining than time evolution. Furthermore the properties of these maps are related to the appearance of (bubble) divergences in spin foams \cite{louapre,smerlak,riello,bonzom}.

\section{Time evolving phase spaces}\label{class}

Here we are going to explain, how to understand discrete time evolution in systems where the phase space dimension (or the `size' of the Hilbert space) can change from one time step to the next. We will consider theories which assume a notion of equal time states, which indeed is the case is many discrete theories, such as Regge calculus \cite{regge} or loop quantum gravity restricted to discrete graphs \cite{ryan,discretelqgreview,spezialeD}. 

First let us illustrate that the need for such a time evolution scheme appears naturally in simplicial discretizations, i.e triangulations. 
 Assume a triangulated hypersurface.  The configuration space of the theory is given by an association of variables to certain type(s) of simplices or combinations of simplices. For instance in (length) Regge calculus \cite{regge} one associates lengths to the edges of the triangulations, other formulations work also with areas \cite{yasha1} or areas and angles \cite{bd09curved}  and references therein. Scalar fields can be associated to vertices, discrete $n$--forms to $n$--simplices or their $n$--duals, etc. \cite{sorkin}.

Time evolution in a $d$--dimensional theory is given by gluing $d$--simplices to the triangulated $(d-1)$--dimensional theory. This discrete time evolution appears as a change of triangulation -- indeed a Pachner move \cite{pachner} -- in the triangulated hypersurface, see also \cite{carfora}.  One can understand Pachner moves as the most elementary change of a triangulation, Pachner moves divide time evolution into basic steps.

'Gluing' a $d$--simplex to the hypersurface, means identifying the variables on the (sub)--simplices that are now shared between this $d$--simplex and the hypersurface as well as solving for (integrating over) the variables that are now in the bulk, i.e. not associated to the hypersurface any more.

The $d$--simplices can be glued to the hypersurface in different ways. Depending on how many faces ($(d-1)$--subsimplices) of a $d$--simplex are identified with $(d-1)$ subsimplices of the triangulated hypersurface, the number of variables associated to the hypersurface might either increase, decrease or stay constant. Accordingly we will interpret these Pachner moves as refining, coarse graining, or of `mixed type'.\footnote{Such moves of `mixed type' can be avoided, if one considers so called Alexander moves instead of Pachner moves. These Alexander moves can be understood as combinations of Pachner moves, thus they will both refine the hypersurface and entangle certain degrees of freedom of this hypersurface.}
 These `mixed type' Pachner moves can be seen as entangling moves, appearing in the entanglement renormalization approach \cite{vidal,koenig}, see the discussion in section \ref{top}.

%%%___________________________________________

%The main point of the paper is to argue that time evolution can be interpreted as refining or coarse graining a given state. In the discrete context this requires a notion of time evolution, which allows a change in the number of degrees of freedom.\footnote{Note that we assume a notion of equal time states here, which indeed is the case is many discrete theories, such as Regge calculus \cite{regge} or loop quantum gravity restricted to discrete graphs \cite{ryan,discretelqgreview,spezialeD}. However, in causal set theory \cite{causal} one might not want to introduce such a notion.} 

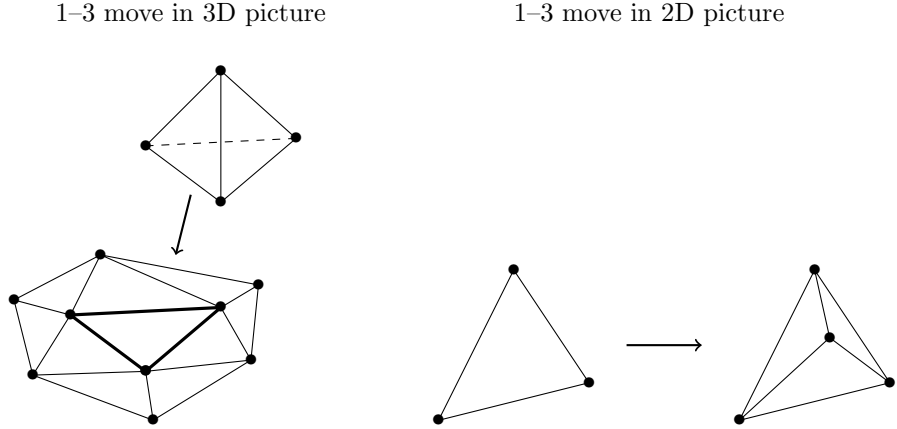
\begin{figure}
\begin{center}
\begin{tikzpicture}
\draw[very thick] (0,0) -- (1,-0.75) -- (2,0.1) -- (0,0);
\draw (0,0) -- (-0.5,-0.8) -- (1,-0.75)
      (0,0) -- (-0.75,0.2) -- (-0.5,-0.8)
      (0,0) -- (0.4, 0.8) -- (-0.75,0.2)
      (0.4,0.8) -- (2,0.1)
      (0.4,0.8) -- (2.5,0.4) -- (2,0.1)
      (2.5,0.4) -- (2.4,-0.6) -- (2,0.1)
      (2.4,-0.6) -- (1,-0.75)
      (2.4,-0.6) -- (1.1,-1.4) -- (1,-0.75)
      (1.1,-1.4) -- (-0.5,-0.8);
\draw (1,2.25) -- (2,1.5) -- (3,2.35);
\draw [dashed] (3,2.35) -- (1,2.25);
\draw (1,2.25) -- (2,3.25) -- (2,1.5)
      (2,3.25) -- (3,2.35);
\draw[->][thick] (1.6,1.6) -- (1.4,0.8);
\draw (0,0) node {\textbullet}
      (1,-0.75) node {\textbullet}
      (2,0.1) node {\textbullet}
      (-0.5,-0.8) node {\textbullet}
      (1,-0.75) node {\textbullet}
      (0.4,0.8) node {\textbullet}
      (2.5,0.4) node {\textbullet}
      (-0.75,0.2) node {\textbullet}
      (2.4,-0.6) node {\textbullet}
      (1.1,-1.4) node {\textbullet}
      (1,2.25) node {\textbullet}
      (2,1.5) node {\textbullet}
      (3,2.35) node {\textbullet}
      (2,3.25) node {\textbullet}
      (1.6,4.0) node {1--3 move in 3D picture};
\end{tikzpicture} \quad \quad \quad
\begin{tikzpicture}
\draw (0,0) -- (2,0.5) -- (1,2) -- (0,0);
\draw (4,0) -- (6,0.5) -- (5,2) -- (4,0)
      (4,0) -- (5.2,1.1) -- (6,0.5)
      (5.2,1.1) -- (5,2);
\draw[->] [thick] (2.5,1.) -- (3.5,1.);
\draw (2.8,5.4) node {1--3 move in 2D picture}
      (0,0) node {\textbullet}
      (2,0.5) node {\textbullet}
      (1,2) node {\textbullet}
      (4,0) node {\textbullet}
      (6,0.5) node {\textbullet}
      (5,2) node {\textbullet}
      (5.2,1.1) node {\textbullet};
\end{tikzpicture}
\caption{\label{31} A $1-3$ move in the 2D hypersurface can be obtained by gluing a tetrahedron with one of its triangles to the hypersurface.}
\end{center}
\end{figure}

For example in $(1+1)$ dimensions, gluing triangles to a triangulated line can be done in two ways, which are named $1-2$ and $2-1$ Pachner move. For the $1-2$ Pachner move we glue a triangle with its base to an edge of the 1D line. This edge is mapped to two edges under time evolution -- which alternatively can be interpreted as refining the state. Indeed we will later see that this is exactly the case in topological theories. In the $2-1$ move we glue a triangle with two edges to two neighbouring edges of the 1D line. 
%This clearly provides one way of interpreting time evolution as coarse graining or refining respectively. Of course one can also devise a time evolution which does not change the number of edges -- for instance by gluing a quadrangle built out of two triangles to the hypersurface. This move can be understood as the basic time evolution move preserving the number of (kinematical) degrees of freedom in a local (that is with only nearest neighbour coupling) way. A non--trivial time evolution operator leads to an entangling (in the quantum theory) of the two degrees of freedom involved, hence such a move can be interpreted as an entangling move.

In 3D one can reproduce the coarse graining $3-1$ and refining $1-3$ Pachner moves by gluing a tetrahedron with three triangles and one triangle respectively, to the  triangulated hypersurface, see figure \ref{31}.

\begin{figure}
\begin{center}
\begin{tikzpicture}
\draw[very thick] (0,0) -- (1,-0.75) -- (2,0.1) -- (1.25,-0.25) -- (0,0)
      (1.25,-0.25) -- (1,-0.75);
\draw (3,-0.6) -- (2,0.1)
      (3,-0.6) -- (1,-0.75)
      (3,-0.6) -- (2.75,0.3) -- (2,0.1)
      (1,-0.75) -- (-0.6,-0.6) -- (0,0)
      (-0.6,-0.6) -- (-0.8,-0.2) -- (0,0)
      (0,0) -- (1.1,0.15) -- (1.25,-0.25)
      (1.1,0.15) -- (2,0.1);
\draw (2,2) -- (3,1.25) -- (4,1.9);
\draw [dashed] (2,2) -- (3.25,1.75) -- (4,1.9)
               (3.25,1.75) -- (3,1.25);
\draw  (2,2) -- (4,1.9);
\draw [<->] [thick] (2.3,1.4) -- (1.4,0.4);
\draw (0,0) node {\textbullet}
      (1,-0.75) node {\textbullet}
      (2,0.1) node {\textbullet}
      (1.25,-0.25) node {\textbullet}
      (3,-0.6) node {\textbullet}
      (-0.6,-0.6) node {\textbullet}
      (-0.8,-0.2) node {\textbullet}
      (1.1,0.15) node {\textbullet}
      (2,2) node {\textbullet}
      (3,1.25) node {\textbullet}
      (4,1.9) node {\textbullet}
      (3.25,1.75) node {\textbullet}
      (2.75,0.3) node {\textbullet}
      (2,3) node {2--2 move in 3D picture};   
\end{tikzpicture} \quad \quad \quad
\begin{tikzpicture}
\draw (0,0) -- (0.5,-0.55) -- (2.0,0.2) -- (1.25,0.6) -- (0,0)
      (1.25,0.6) -- (0.5,-0.55);
\draw (4,0) -- (4.5,-0.55) -- (6.0,0.2) -- (5.25,0.6) -- (4,0)
      (4,0) -- (6,0.2);
\draw [<->][thick] (2.5,0) -- (3.5,0);
\draw (0,0) node {\textbullet}
      (0.5,-0.55) node {\textbullet}
      (2.0,0.2) node {\textbullet}
      (1.25,0.6) node {\textbullet}
      (4,0) node {\textbullet}
      (4.5,-0.55) node {\textbullet}
      (6.0,0.2) node {\textbullet}
      (5.25,0.6) node {\textbullet}
      (2.5,3.2) node {2--2 move in 2D picture};
\end{tikzpicture}
\caption{ \label{22} A $2-2$ move in the 2D hypersurface can be obtained by gluing a tetrahedron with two of its triangles to the hypersurface.}
\end{center}
\end{figure}
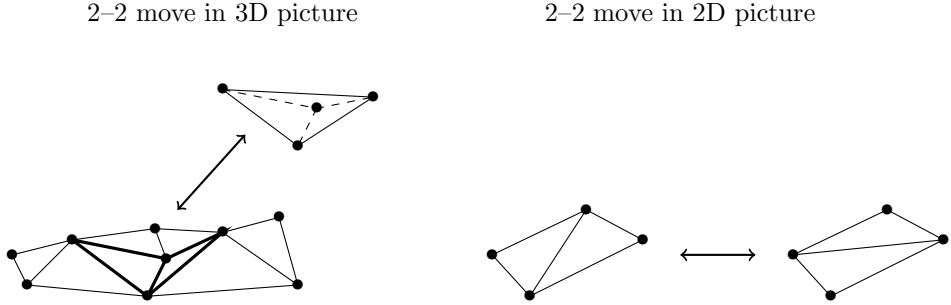

However if one wants to produce a very refined state and uses only $1-3$ Pachner moves one will end up with a very peculiar geometry, known as stacked sphere. Even in 4D, where gravity is non--topological and interacting,  such stacked sphere geometries are not dynamical (do not allow for curvature) and span the flat sector of the theory as defined in \cite{ryan}.
%\footnote{Such stacked spheres play also an essentially role in dynamical triangulations, where they dominate the weak coupling phase of simplicial gravity \cite{edt}. This phase reappeared as melonic phase in the more modern re--incarnation of dynamical triangulations as coloured tensor models \cite{gurau}. 
%However also there one can argue that the encoded geometries are degenerate, leading to lower Hausdorff dimensions \cite{gurauryan}.} 
%
Thus, to arrive at more interesting spatial geometries one needs to include other Pachner moves. For $(2+1)$ dimensions these are the $2-2$ moves which can also be obtained by gluing a tetrahedron with two triangles to the hypersurface, see figure \ref{22}. Such $2-2$ moves can be used as entangling moves  to produce the long range entanglement in topological phases  \cite{koenig}. %(i.e. $(2+1)$ gravity with a positive cosmological constant that can be described by a Tuarev-Viro model \cite{tuarev-viro} and give  string net models in condensed matter \cite{wenstringnet}). 
For $(3+1)$ dimensions one can generate analogously $4-1$ and $1-4$ as well as $3-2$ and $2-3$ Pachner moves by gluing a 4--simplex to the 3D triangulated hypersurface.

These moves can be described via canonical evolution equations, despite the change in phase space dimension \cite{hoehn1,hoehn2,hoehn3}.
Generalizing work of \cite{marsden,gambini,bahrreview} such discrete time evolution maps can be understood as canonical transformations generated by an action.
This action is associated to the $d$--simplices and can be understood as Hamilton's principal function depending on the boundary data of this simplex.\footnote{The advantage of such a formulation is that it reflects how simplicial path integrals are defined. There, i.e. in spin foams, one associates an amplitude to a $d$--simplex, which in the semi--classical limit does indeed give the Regge action for the simplex \cite{asympt}.  The path integral (say with boundary) is then defined by summing the product of all simplex amplitudes over all bulk variables. Thus changing the boundary state by gluing a simplex  to the boundary (i.e. multiplying the state with the simplex amplitude and summing over the variables which are now bulk variables), we automatically obtain the amplitude for the evolved state.  Hence one would expect that the semi--classical limit reproduces the equation of motion as obtained from the canonical transformation generated by the action associated to this simplex.} Hamilton's principal function is a generating function for the momenta, that is we use the action associated to a simplex $S_s$ to define old momenta $p$ and new momenta $p'$. Schematically we have
\ba\label{canevol}
p\,=\,&\frac{\partial S(q,q')}{\partial q}  \q ,\q\q
p'\,=\,-\frac{\partial S(q,q')}{\partial q} \quad ,
\ea
where we denote old and new configuration data by $q$ and $q'$ respectively.\footnote{Some configuration variables are neither old or new, as these are represented in the hypersurface before and after the move. Such variables count as (non--dynamical) parameters in this move. Here we will only need this schematic discussion, for explicit discussion of all Pachner moves see \cite{hoehn2}.} 

How can the equations (\ref{canevol}) describe a canonical, i.e. symplectic, transformation, if the number of old and new variables differ? The answer is that pre-- and/or post-- constraints appear on the initial or final phase space respectively. One has to reduce the phase spaces with respect to these constraints and finds a symplectic transformation on these reduced phase spaces.

The constraints have to appear for a simple reason from the equations (\ref{canevol}). There one would have to solve the first set of equations for the new configurations in terms of the old configurations and old momenta. However, if we have $N_{old}>N_{new}$ variables, the first set of equations will give $N_{old}$ relations for $N_{new}$ unknowns. Thus, if the equations are independent, they will give the solutions $q'(q,p)$ but also  $(N_{old}-N_{new})$ pre--constraints $C_i(q,p),i=1,\ldots,(N_{old}-N_{new}) ,$ that is constraints on the initial phase space. Similarly we obtain post--constraints, if $N_{new}>N_{old}$. (Constraints can also appear independently of this mechanism, that is $N_{new}=N_{old}$ does not guarantee that constraints will not appear.) 
As the pre-- or post--constraints are defined via a generating function, they are first class. Thus the evolution equations (\ref{canevol}) leave a number of configurations undetermined ('pre--and post gauge degrees of freedom), corresponding to the number of constraints that appear. The status of these gauge degrees of freedom might change under further evolution: constraints appearing in the future might lead to a gauge fixing. 
%Again one might wonder how a canonical transformation is possible between phase spaces of different dimensions. The answer is to enlarge (if needed) both phase spaces, the initial and the final one, so that the kinematical degrees of freedom can be matched to each other. Secondly the action has to be interpreted in the right way as a generating function of the `old' and `new' variables. This allows to obtain the canonical evolution in the standard way from a generating function, namely the action. The artificial enlargement of the phase spaces is reflected in the appearance  of pre-- and post--constraints, which will be essential for our interpretation of time evolution as refining and coarse graining maps. 
The pre--constraints have to be satisfied for an evolution move to take place. The post--constraints are automatically satisfied, after an evolution move has taken place.

We should point out that the pre-- and post--constraints include constraints which might arise due to gauge symmetries, including Hamiltonian and diffeomorphism constraints. 
For instance the $4-1$ Pachner move in 4D  leads to Hamiltonian and diffeomorphism constraints \cite{ryan,hoehn2}. In this case the post--constraints exactly coincide with the Hamiltonian and diffeomorphism constraints, no new truly physical degree of freedom is added by such a refinement move. There are however also the $2-3$ moves that add degrees of freedom and therefore lead to constraints, which do however not coincide with the Hamiltonian and diffeomorphism constraints. 
As noted further evolution might fix the gauge degrees of freedom implied by the Hamiltonian and diffeomorphism constraints. This is due to the breaking of diffeomorphism symmetry in discretization of 4D gravity \cite{bahrdittrich09a}
%, at least for 4D gravity, such constraints might present themselves rather as pseudo--constraints \cite{gambini,bahrdittrich09a}.  

Post-- and pre-- constraints also appear for theories without any a priori gauge symmetries, such as a scalar field theory. We propose here that such constraints can be interpreted as describing the state of finer degrees of freedom. We will motivate this proposal with examples.

\subsection{ Example: evolution of a scalar field on an extending triangulation}\label{subscalar}

\begin{figure}
\begin{center}
\begin{tikzpicture}[scale=0.8]
\draw [dashed] (-4,-1.5) -- (-4,1.5)
               (4,-1.5) -- (4,1.5);
\draw (-4,-1.5) -- (4,-1.5)
      (-4,1.5) -- (4,1.5);
\draw (-4,1.5) -- (-3.1,-1.5) -- (-2.4,1.5) -- (-1.1,-1.5) -- (-0,1.5) -- (0.8,-1.5) -- (2.3,1.5) -- (3.5,-1.5) -- (4,1.5);
\draw (-3.2,1.5) -- (-3.1,-1.5)
      (-1.5,1.5) -- (-1.1,-1.5)
      (1.2,1.5) -- (0.8,-1.5)
      (3.0,1.5) -- (3.5,-1.5);
\draw (-3.1,-1.75) node {$1$}
      (-3.1,-1.5) node {\textbullet}
      (-1.1,-1.75) node {$2$}
      (-1.1,-1.5) node {\textbullet}
      (0.8,-1.75) node {$3$}
      (0.8,-1.5) node {\textbullet}
      (3.5,-1.75) node {$4$}
      (3.5,-1.5) node {\textbullet}
      (-4,1.75) node {$1'$}
      (-4,1.5) node {\textbullet}
      (-3.2,1.75) node {$2'$}
      (-3.2,1.5) node {\textbullet}
      (-2.4,1.75) node {$3'$}
      (-2.4,1.5) node {\textbullet}
      (-1.5,1.75) node {$4'$}
      (-1.5,1.5) node {\textbullet}
      (0,1.75) node {$5'$}
      (0,1.5) node {\textbullet}
      (1.2,1.75) node {$6'$}
      (1.2,1.5) node {\textbullet}
      (2.3,1.75) node {$7'$}
      (2.3,1.5) node {\textbullet}
      (3.0,1.75) node {$8'$}
      (3.0,1.5) node {\textbullet}
      (4.0,1.75) node {$1'$}
      (4.0,1.5) node {\textbullet};             
\end{tikzpicture}
\caption{The scalar field on a circle and its time evolution. The circle is drawn as an interval with periodic boundary conditions indicated by the dashed lines. \label{fig3}}
\end{center}
\end{figure}
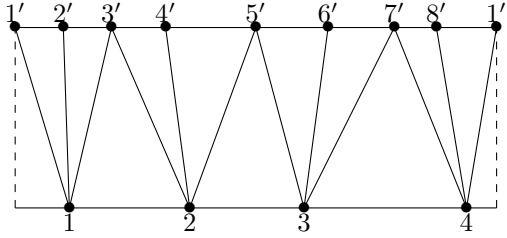

As a first  example  we consider a massless scalar field on a 2D (Euclidean) equilateral triangulation. The action associated to one triangle is given as
\ba\label{triang}
S_\Delta &=& \frac{1}{4}\sum_{e\subset \Delta} (\phi_{s(e)}-\phi_{t(e)})^2 \quad ,
\ea
where $s(e),t(e)$ denote the source and target vertex of an oriented edge respectively. We now consider a time evolution between two spatial periodically identified 1D triangulations, i.e. circles subdivided into edges. We assume that the earlier equal time hypersurface has $N$ edges and the later one $N'$ edges and we connect these two hypersurfaces by ``one slice of triangles'', see figure \ref{fig3}. The triangulation can be described by an adjacency matrix $A_{vv'}$, where $A_{vv'}=1$ if the vertex $v$ at the earlier time is connected to the vertex $v'$ at the later time, and $A_{vv'}=0$ if this is not the case. The canonical time evolution map can be easily computed in this case. In particular the momenta $\pi'_{v'}$ at the later time step are given by
\ba\label{ex1}
\pi'_{v'}&=&\frac{\partial S}{\partial \phi_v}\,=\, \left ( \sum_v A_{vv'} (\phi'_{v'} - \phi_v)\right) + \phi'_{v'}-\tfrac{1}{2}\phi'_{v'+1} -  \tfrac{1}{2}\phi'_{v'-1} \quad ,
\ea
where $S$ is the action associated to the interpolating triangulation obtained by summing the action contributions (\ref{triang}) of the triangles.  If $A_{vv'}$ has right null vectors $R^{v'}_r$, i.e. such that $A_{vv'}R^{v'}_r=0$, we obtain constraints by contracting (\ref{ex1}) with these null vectors. These constraints are of the form
\ba\label{ex2}
C_r &=&  \sum_{v'}R^{v'}_r \pi'_{v'} \,\,+\,\, f_r(\phi') 
\ea
with some specific functions $f_r$ of the fields $\phi'_{v'}$ at the later time step. Coming from a generating function the constraints are Abelian. They generate gauge transformations in the sense that the evolution step leaves indeed certain combinations of field values unspecified. Assuming an orthogonal basis of the null vectors, these combinations are given as $\lambda_r=\sum_{v'} R^{v'}_r \psi'_{v'}$. For a refining evolution step we have a larger number $N'$ of vertices at the later time  than the number of vertices $N$ at the earlier time, $N'>N$. In this case there are at least $N'-N$ right null vectors $R_r$, with $r=1,\ldots, N'-N$. We want to argue that these gauge degrees of freedom correspond to the finer degrees of the field.

Consider specifically a regular refinement as in figure \ref{fig3}, with $N'=2N$. The corresponding adjacency matrix is given by
\ba\label{ex3}
A_{vv'}&=& \delta_{2v-1,v'}+\delta_{2v,v'}+\delta_{2v+1,v'} \q .
\ea
The matrix can be `diagonalized' by a Fourier transform. Let us formally introduce the notation
\ba
& |k\rangle\,=\, \sum_{v=1}^N e^{-2\pi i \frac{kv}{N}} |v\rangle  \q\q, \q & |v\rangle\,=\, \frac{1}{N}\sum_{k=0}^{N-1} e^{2\pi i \frac{kv}{N}} |k\rangle \nn\\
& |k'\rangle\,=\, \sum_{v'=1}^{2N} e^{-2\pi i \frac{k'v'}{2N}} |v'\rangle  \q\q, \q & |v'\rangle\,=\, \frac{1}{2N}\sum_{k'=0}^{2N-1} e^{2\pi i \frac{k'v'}{2N}} |k'\rangle \quad ,
\ea
so that we can write
\ba\label{ex4}
\langle k|A|k'\rangle ,=\,\sum_{vv'}\langle k|v\rangle A_{vv'}\langle v'  |k'\rangle&=& \left(1+ e^{\pi i \frac{k'}{N}} + e^{-\pi i \frac{k'}{N}} \right) \delta^{(N)}(k,k') \nn\\ &=&\left(1+2\cos \left(\frac{k\pi}{N} \right)\right) \delta(k,k') + \left(1-2\cos \left(\frac{k\pi}{N} \right)\right) \delta(k+N,k') \q .\q\q
\ea
We thus have $N$ right null vectors $R^{k'}_k$ (here $k$ labels the null vectors) given by 
\ba
\sum_{k'}R^{k'}_k |k'\rangle&:=&     \left(1-2\cos \left(\frac{k\pi}{N} \right)\right) |k\rangle  \,\,-\,\,  \left(1+2\cos \left(\frac{k\pi}{N}\right)\right) |k+N\rangle \quad .
\ea
In general the coefficient in front of the higher momentum $|k+N\rangle$ is non--vanishing (it only vanishes for $k=2N/3$). Thus we can say that (almost) all  momenta $\pi'_{k'}$ associated to finer degrees of freedom, i.e. with momenta $k'>N$ are determined by the constraints (\ref{ex2}).  The same holds if we add a potential $V$ to the scalar field, and discretize this in a local manner, i.e. as a term $\sum_{v \in \Delta}  Ar(\Delta,v) \,V(\phi_v)$  added to the action (\ref{triang}) with $Ar(\Delta,v)$ denoting some association of an area to the vertex--triangle pair. 

~\\

The post--constraints signify in particular that no new information is added, the physical phase space cannot be enlarged during evolution.\footnote{For a complete discussion on how these constraints propagate and a classification of the constraints, see \cite{hoehn3}.} In fact, we can interpret this in the following way: given a state with a certain coarse graining, i.e. discretization scale, we can apply refining time evolution steps.  This will lead to a state with the same coarse graining scale, however represented on a finer discretization. 

It is preferable that the finer degrees of freedom that are added during refining time evolution are in a vacuum state. In the case of a scalar field we have a notion of energy, thus the statement is that the refining time evolution should not increase the energy as defined by some energy functional on the two different discretizations. See also the discussion in \cite{foster}, which considers this issue however in a covariant quantization scheme. Whether this is actually the case will depend on the quality of the discretization.

This is particularly true because the degrees of freedom that are added are typically defined on scales near the discretization scale. Typical discretizations will give unreliable results on this scale. A way out is to design discretizations so that the added degrees of freedom are in fact in a vacuum state.

\subsection{Refinement as adding degrees of freedom in the vacuum state}\label{refv}

In the last section \ref{subscalar} we have used the Fourier transform to identify finer degrees of freedom (higher modes) and coarser degrees of freedom (lower modes). In fact in a free theory the Fourier modes decouple and allow us to assign an energy per mode.

We can thus make the argument that refining time evolution should add degrees of freedom in a vacuum state and design a discretization for which this is the case. This will in general result in a non--local (in space) discretization -- as can be already suspected if one uses the Fourier transform.

The general idea is to match for a set of Fourier modes up to a cut--off $K$ exactly the dynamics of the continuum, see also the related arguments in \cite{kempf,unpublished}.

The construction is as follows:
We consider the canonical data of a  continuum scalar field on a 1D circle, representing the spatial hypersurface. We only consider fields that include Fourier modes $\tilde \phi(k_1)$ up to a cut-off $K \in {\mathbb N}$ on the spatial momentum component $|k_1| \leq K$, so that the fields are superpositions of $N=2K+1$ modes. 
Such fields can  therefore be parametrized  one--to--one (in the general case) by the set of $N$ values of the field at pre--specified positions along the circle.  

We thus have a mapping ${\cal M}_{K}$ from the phase space describing continuum field configurations with a cut-off $K$ to a phase space describing a discretized scalar field on $N=2K+1$ vertices. 

We have now to decide on an embedding map ${\cal C}_{K,K'}$ from the phase space with cut--off $K$ to a phase space with $K'\geq K$. Once such a map is chosen we can construct the corresponding map ${\cal D}_{K,K'}={\cal M}_{K'} \circ  {\cal C}_{K,K'}  \circ ({\cal M}_{K})^{-1}$ for the phase spaces describing the discretized fields.

As an example, one can choose ${\cal C}_{K,K'}$ such that in a mode expansion the coefficients of the additional modes are vanishing. This minimizes the energy of the additional modes for a free theory. For interacting theories one can choose more generally ${\cal E}_{K,K'}$ such that the energy of the refined configurations (in the space of fields with a mode cut-off $K'$) is minimized, keeping the coarser modes fixed.

Note also that one can attempt to define an embedding  ${\cal C}_{K,K'}$ such that it includes some proper time evolution step ${\cal T}$.
However a time evolution will entangle the modes up to the cut-off $K$, with (possibly) all continuum degrees of freedom, that is the image of ${\cal T}$ applied to a phase space given by modes with cut--off $K$ will in general include modes $K' \rightarrow \infty$.
We therefore face a problem in pulling back the evolved continuum configuration to a discrete one,  as the evolved continuum configuration might be infinitely refined with respect to the embedding maps chosen.

Thus one must find an (embedding) map of discrete configurations into continuous ones, where this is not the case. In this sense the choice of the (generalized) maps ${\cal M}_K$ should be informed by the dynamics. We believe however that such examples are rare, see section \ref{scalar} for one of these cases. 
Alternatively, one chooses a truncation of the image of ${\cal T}$ back to the phase space describing modes with a cut--off $K'$. 
This introduces an approximation to the continuum dynamics. However for sufficiently refined configurations, one would of course expect, that these errors do not affect sufficiently coarse grained observables.

In general the discrete embedding maps ${\cal D}_{K,K'}$ will be highly non--local, but this will be necessary to obtain a good approximation also for modes near the discretization scale. This construction will be very involved for an interacting theory, as it basically requires the solution of the dynamics.

However it can serve as a guide line of what to expect from `good' discretizations, that also involve a possible change of degrees of freedom. Thus even if one has a discretization that does not exactly mirror this behaviour (i.e. is not `perfect') one can hope that via coarse graining one reaches an effective theory, that actually does so. 
This is the philosophy behind   perfect discretization, which can be constructed as fixed points of renormalization flows \cite{hasenfratz, bahrdittrich09b,he}  or by pulling back continuum physics to the lattice (`blocking from the continuum') \cite{bietenholz}. A more abstract approach is to select as observables spectra of geometric operators \cite{kempf2}.

The construction described in this section allows a more explicit choice of the (post)--constraints, than in the discussion in section \ref{class}, where the post--constraints are determined by the chosen discretization of the action.
 Here the post--constraints are determined by the choice of vacuum state, given by the minimization of an energy functional. Note that the constraints are second class. For instance for a free theory these are given by the vanishing of all higher modes in the fields and momenta. Thus in comparison with the discussion in section \ref{class} one has gauge fixed the first class post--constraints appearing there with additional constraints.

This is what one would however expect also from the quantization of a (free) scalar field: the vacuum functional in a given mode is given as a Gaussian of the field variable. Such a  Gaussian can also be found by minimizing the `master constraint' $M(k_1)= \omega^2(k) \tilde \phi^2 (k_1)+  \tilde \pi(k_1)$, given as sum of (weighted) squares of the individual constraints \cite{master}. For gravity the situation is less clear what kind of vacuum to expect. On the one hand (continuum) gravity constitutes a first class constrained system, so all physical states have to satisfy these constraints. Thus, physical states are squeezed states in the conjugated degrees of freedom describing the gauge choice and the constraints. In 4D we of course have additional physical degrees of freedom, however the characterization of a vacuum state (without a background and boundary) is an open issue. We will discuss in section \ref{timeevol} the Hartle Hawking no--boundary proposal \cite{hhv} for a vacuum state, that can be naturally implemented with a refining time evolution.

\subsection{Massless scalar field in a 2D Lorentzian space time}\label{scalar}

Here we will discuss an example of a perfect discretization with local embedding maps, namely the discretization of a massless scalar field in 2D Minkowskian space time. Note that this is the only such example of a non--topological theory that we are aware of, and that the locality of the embedding maps might actually change in the quantum theory. % \cite{bdtoappear}.
 We will consider  equal time hypersurfaces  given by piecewise null lines, akin to characteristic evolution schemes \cite{lehner}. 

We will identify a given discrete configuration of field values with a continuous configuration by assuming the continuum field to be piecewise linear. Such a piecewise linear field can be parametrized by a discrete set of scalar field values at points where the derivative of continuum field is not continuous.

One motivation for this example is to provide an interpretation for graph changing Hamiltonians appearing in loop quantum gravity \cite{thiemann, madha} or for the parametrized scalar field \cite{madhavan}. This example will illustrate that indeed refining  evolution splits into an embedding map and a proper evolution. 

%We will introduce an example, the 2D scalar field on Minkowskian space time. This example generalizes to conformally flat (hence all) 2D Lorentzian space times (on a cylinder).  It allows -- at least on the classical level -- a description of the discrete perfect dynamics in an (almost) local way. This might actually change in quantum theory \cite{bdtoappear}. It also is peculiar to the Lorentzian signature and the massless case. Indeed, it is the only example with propagating degrees of freedom, that we know of, for which this is the case.

The example is furthermore interesting as it introduces the concept of piecewise null hypersurfaces, that on `larger scales' can be either put together to a spatial hypersurface, or alternatively to a null hypersurface. Thus problems involving a null boundary can be easily treated, with a natural specification of `boundary conditions' at the null hypersurface (or null line). %This will be explored in the forthcoming work \cite{bdtoappear}.  
For a discussion of issues related to holography involving such discretizations, see \cite{yasha}, which very much inspired the development of this example. Null surface formulations also attracted recent interest in (loop) gravity \cite{reisi}.

To be concrete we consider a 2D cylinder %\footnote{The circumference of the cylinder will introduce a scale in the system.} 
space time endowed with the Minkowski metric. We consider piecewise null `hyperlines' that close around the cylinder. Thus we will have null edges connected via kinks.

For every such kink we have to introduce a vertex $\nu$. We allow furthermore vertices $\nu$ on the null edges themselves. We will associate scalar field values $\phi_\nu$  to these vertices $\nu$.

As we will show in the following, such a configuration of scalar fields $\phi_\nu$  specifies a piecewise linear solution to the continuum dynamics.

Let us start with the set of continuum solutions to $\square \phi=0$, which are given by
\ba\label{sol}
\phi(u,v)=f(u)+g(v)
\ea
where $(u=t+x,v=t-x)$ are light cone coordinates\footnote{Choosing $x\in [0,2\pi)$ we have $\tfrac{1}{2}(u-v)\in [0,2\pi)$.}. We will consider functions of the form (\ref{sol}) with $f$ and $g$ piecewise linear (continuous) functions.  Thus $\phi(u,v)$ will be smooth (even linear) everywhere except at a  set of null lines $u=c_I$ or $v=c'_J$, where $\{c_I,c'_J\}$ are a set of constants.  Such a solution induces a scalar field configuration on any piecewise null line in the following way:  

As outlined above, we have a vertex at every kink of the piecewise null line. Additionally we introduce vertices for every null line $u=c_I$ or $v=c'_J$ that cuts our `equal time hypersurface' transversally. The values of the scalar field at these vertices are now just given by the values of the solution (\ref{sol}) at the position of the vertices.

This gives a configuration of scalar field values on a piecewise null hyperline. From this configuration we can re--construct the solution. We basically do the inverse of the above procedure: For every kink we draw two null lines $u=c_I$ and $v=c'_J$  emanating from this kink. Furthermore we draw from every vertex on a null edge a transversal null line. These null lines give the possible non--smooth behaviour of the solution. We can reconstruct the solution everywhere by linear extrapolation.

A concrete way of constructing such a solution is given by a time evolution of the scalar field configuration on a piecewise null line, by pushing the null line forward in time.  %We will here describe a time evolution that moves the null edges forward in time. 

Note that the scalar field values on a given piecewise null line  are sufficient to reconstruct the full space time solution. There are no additional momenta needed. Intuitively this can be imagined the following way: drawing a null zigzag line we obtain a set of initial fields at two consecutive time steps. Thus the fields themselves provide the momenta. %Indeed one can adapt a symplectic structure from the continuum, in which this is exactly the case \cite{bdtoappear}. 

\begin{figure}
\begin{center}
\begin{tikzpicture}
\draw (0,0) -- (1,1) -- (2.5,0) -- (3,0.5) -- (4,1.5) -- (4.5,2);
\draw [dashed] (1,1) -- (1.5,1.5) -- (3,0.5);
\draw (0.75,1.25) node {$\nu_1$}
      (1,1) node {\textbullet}
      (2.5,-0.25) node {$\nu_2$}
      (2.5,0) node {\textbullet}
      (3,0.5) node {\textbullet}
      (3.25,0.25) node {$\nu_2'$}
      (1.5,1.5) node {\textbullet}
      (1.25,1.75) node {$\nu_1'$}
      (4,1.5) node {\textbullet}
      (4.25,1.25) node {$\nu_3$}
      (1.4,1.15) node {$\epsilon$};
\end{tikzpicture}
\caption{\label{null1} The time evolution proceeds by moving the null edge $(\nu_1,\nu_2)$ to $(\nu'_1,\nu'_2)$.}
\end{center}
\end{figure}
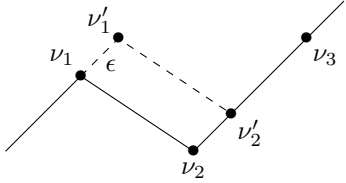

To describe the time evolution consider a piecewise null line with a vertex $\nu_1$ at a kink. We wish to evolve this vertex by an amount $\epsilon$ in say the direction of $u$, see figure \ref{null1}.  (Note that there is no absolute length attached to $\epsilon$ as it is an affine parameter. The time evolution itself can only be characterized by the area of the rectangular diamond that will be glued to the hypersurface.)  Let us assume (for simplicity) that the next vertex $\nu_2 $ to the right of $\nu_1$ is also a kink. If we want to move the vertex $\nu_1$ and keep the hypersurface null we have to also move the vertex $\nu_2$ to a new vertex $\nu'_2$. Thus we move an entire null edge of the hyperline.

Note however that although we move the vertex $\nu_1$ to a new vertex $\nu'_1$ we actually have to keep the vertex $\nu_1$ as a vertex in our hypersurface. This is due to the possible non--smooth behaviour in the field that might still occur at $\nu_1$. Thus the new hypersurface will have one additional vertex. In this way time evolution is necessarily\footnote{One can also time evolve by gluing diamonds that fit into the zigzag null line. This would keep the number of vertices constant, but would also pre--define the size of the time evolution step.} refining. 

Thus we have to determine the values of the scalar field at the new vertices $\nu'_1$ and $\nu'_2$. We construct the field $\phi(\nu'_2)$ by linearly interpolating between $\phi(\nu_2)$ and the field $\phi(\nu_3)$ at the next vertex $\nu_3$ to the right of $\nu_2$:
\ba\label{f1}
\phi(\nu'_2) =  \phi(\nu_2) +   \frac{\phi(\nu_3)-\phi(\nu_2)}{u(\nu_3) -u(\nu_2)}(u(\nu'_2)-u(\nu_2))  \q .
\ea
(One might consider this step to be somewhat non--local.) Having constructed the field $\phi(\nu'_2)$ we now know three fields at the vertices of the diamond formed by $\nu_1,\nu'_1,\nu_2,\nu'_2$. The field at the tip $\nu'_1$ of this diamond is imposed by the form of the solution (\ref{sol}) to be  
\ba
\phi(\nu'_1)= \phi(\nu_1)\,+\, \phi(\nu'_2) \,-\,\phi(\nu_2) \q .
\ea

This description can be easily generalized to other situations. In the end time evolution proceeds by gluing rectangular diamonds to the null hypersurface.  Allowing the side length of these diamonds to vary, we do not pick out a Lorentz frame, thus we have a Lorentz independent cut--off. 

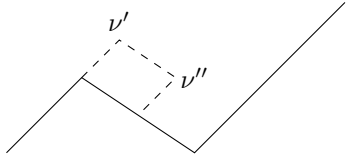
\begin{figure}
\begin{center}
\begin{tikzpicture}
\draw (0,0) -- (1,1) -- (2.5,0) -- (3,0.5) -- (4,1.5) -- (4.5,2);
\draw [dashed] (1,1) -- (1.5,1.5) -- (2.25,1) -- (1.75,0.5);
\draw (1.5,1.75) node {$\nu'$}
      (2.5,1) node {$\nu''$};
\end{tikzpicture}
\caption{ \label{null2} Time evolution can also proceed by gluing small diamonds to the hypersurface, which will however produce past directed null edges. Note that in this case $\phi(\nu'')$ will be constrained and determined by the fields at the other vertices on the `equal time' hypersurface.}
\end{center}
\end{figure}

 Here we have a situation very similar to loop quantum gravity \cite{thiemann}, or parametrized and polymerized scalar field theory \cite{madhavan}, with a `graph changing' Hamiltonian. One can choose $\epsilon$ (or the area of the diamond if one generalizes the framework to allow also 'past directed' null edges, see figure \ref{null2})  arbitrarily small -- there remains a discontinuous action of the time evolution, which is to produce new vertices.

The time evolution map can be also split into two parts: one is a pure refining part, introducing the vertex $\nu'_2$ and the associated field value (\ref{f1}). The second part is the `proper' time evolution step, in which a diamond with vertices $\nu_1,\nu'_1,\nu_2,\nu'_2$ is glued to the hypersurface. This part keeps the number of vertices constant, as the new vertex $\nu'_1$ is compensated by the loss of the old vertex $\nu_2$.

The scheme we have described is a perfect discretization, that exactly mirrors the continuum solutions. We can also embed any discrete configuration into a refined configuration, where the new field values are given by linear interpolation as in (\ref{f1}). This allows to identify discrete and continuum configurations, which can be formalized into an inductive limit construction, which we will explain in section \ref{emb} for the quantum theory.  See \cite{ziprick1} for an alternative proposal to identify discrete and continuous phase space configurations, based on gauge fixing infinitely many degrees of freedom of the continuum theory.
%A discussion of the quantum theory of this system will appear in \cite{bdtoappear}. 

\subsection{ Refinement moves in simplicial gravity: adding gauge and vacuum degrees of freedom}

The examples discussed here involved  scalar field theories, for which a notion of energy and hence vacuum is available. In section \ref{subscalar} we proposed to use directly an energy functional to characterize the state of the additional degrees of freedom. For the massless scalar field on null lines we determine the values of additional `finer'  fields by demanding piecewise linearity of the field. 

In case of gravitational theories the notion of energy is less clear, in particular if one considers compact spatial slices.\footnote{ However one can attempt to proceed similar to the examples in the previous sections: For instance for simplicial (Regge) gravity one can construct, similar to the massless field, a refining based on piecewise flat geometry, at least on the classical level. Another possibility is to use some quasi-local notion of energy and to minimize this energy for the region that is being refined, in analogy to the procedure described in section \ref{subscalar}.} 
 Time evolution itself is rather understood as a gauge transformation and energy is constrained to vanish (on space times with compact spatial slices). Indeed as mentioned in section \ref{class}, part of the degrees of freedom added in a refining time evolution will be gauge (or pseudo gauge if diffeomorphism symmetry is broken \cite{gambini, bahrdittrich09a}). But in general refining time evolution  also adds physical (non--gauge) degrees of freedom. Thus refining time evolution leads to states that are supposed to be gauge equivalent, but seem to be based on different number of degrees of freedom. 

To resolve this puzzle, we need to understand states resulting from a refining time evolution as equivalent -- they represent the same state on different discretizations.  We will  discuss quantum theory in section \ref{emb} in which this notion can be indeed made precise.
For this interpretation it is important that the refined degrees of freedom are indeed in a state, that can be interpreted as vacuum. For instance in loop quantum gravity one uses the so--called Ashtekar--Lewandowski vacuum \cite{al}, in which spatial geometry is sharply peaked to be totally degenerate.
An alternative vacuum state has been recently introduced \cite{geiller}, in which the vacuum is rather peaked on flat connections. In both cases these vacua are used to define a notion of refining, in the second case this refining origins indeed from a time evolution of $BF$ theory, a topological theory that describes flat connections.

However both these choices are rather kinematical vacua, at least in 4D gravity (\cite{geiller} gives actually the physical vacuum for 3D gravity). The interpretation as vacua is not tied to an energy functional, but rather to the fact that these states are the simplest possible ones from different viewpoints. The Ashtekar--Lewandowski vacuum can be understood as the state giving a constant value to all connection fields, whereas the $BF$ vacuum \cite{geiller} gives a constant value to the conjugated variables, the flux fields describing spatial geometry.

Thus, although these vacua are not physical, they show an important property of vacua (in homogeneous systems), namely to be homogeneous. We expect a physical vacuum to be given by a state satisfying the constraints and carrying a notion of homogeneity.  Again the peculiarities of general relativity make this description of homogeneity non--trivial. Physical observables have to be invariant under space--time diffeomorphisms. To nevertheless allow for local observables one can use relation observables \cite{partial} , that often use a reference system built from matter (fields). Such a reference system requires, however, an inhomogeneity in these matter fields, which are used as rods and clocks. An alternative are reference systems built out of gravitational degrees of freedom, such as in \cite{perturb}. There physical observables are constructed that describe perturbations away from homogeneity, thus a vacuum state can be described via a prescription for the expectation values and fluctuations of these observables. 

A different characterization of vacuum uses the notion of path integral as a projector on physical states. Assuming one can construct such a (consistent) projector we can define the image of any of the kinematical vacua under this projector as a physical vacuum. The kinematical vacua are homogeneous states -- therefore one would expect the physical vacuum obtained by projection (assuming it can be constructed) also to be homogeneous.
A priori it is not clear whether e.g. the Ashtekar--Lewandowski vacuum \cite{al} and the  vacuum, based on $BF$ theory \cite{geiller} would lead to the same physical vacuum. As we will describe later, if we construct the projector on physical states via a refining time evolution operator (i.e. a path integral) such vacua can be also understood to realize the  Hartle--Hawking no--boundary proposal for a vacuum \cite{hhv}.

\section{Refining in quantum theory }\label{quant}

Here we will discuss some aspects of refining time evolution in quantum theory. See also \cite{hoehnq}, which introduces a framework for time evolving Hilbert spaces, in which the number of degrees of freedom can increase and decrease. In this work we will base our discussion more on inductive limit Hilbert spaces and rather see refining time evolution as a means to define embedding maps needed for the construction of these inductive limit Hilbert spaces. We will explain  this construction shortly in section \ref{emb}.

%There are two main aspects we will consider: One is that there is a framework available, which implements Hilbert spaces associated to different discretizations. In this framework -- which is used in loop quantum gravity, one defines a continuum Hilbert space  as inductive limit of a family of Hilbert spaces associated to discretizations. We will explain shortly this construction.

So far, this framework has been used on the kinematical level in loop quantum gravity. The main proposal of this work is that one should actually use the dynamics, that is time evolution, to define the embedding maps needed for this framework. 
With this proposal one adopts a no--boundary Hartle Hawking state as vacuum state, thus the vacuum (and a notion of equivalence between states of different refinement degree) is determined by the dynamics of the theory. This will be lined out in section \ref{timeevol}.

\subsection{Inductive limit construction of a continuum Hilbert space}\label{emb}

The inductive limit construction allows to define a continuum Hilbert space from a family of Hilbert spaces associated to discretizations (for instance graphs as in the Ashtekar Lewandowski representation \cite{al} or triangulations as in the $BF$ vacuum introduced in \cite{geiller}). The discretizations need to be organized into a directed partially ordered set, denoted by $(\{b\},\prec)$. The ordering provides a notion of coarser and finer discretizations, that is $b\prec b'$ denotes that $b'$ is a refinement of $b$. In a directed partially ordered set one can always find a common refinement $b''$ for two discretizations $b$ and $b'$.

 We associate to each such discretization $b$ a Hilbert space of states ${\cal H}_b$. For any two Hilbert spaces ${\cal H}_b$ and ${\cal H}_{b'}$ with $b\prec b'$,  we need to define an embedding map
\ba
\iota_{bb'}: {\cal H}_b \rightarrow {\cal H}_{b'} \q .
\ea
These embedding maps have to satisfy consistency conditions: For any $b\prec b' \prec b''$ we demand
\ba\label{cons1}
\iota_{b'b''} \circ \iota_{b b'} &=& \iota_{bb''}  \q.
\ea
As we will see, these conditions encode, under the identification of the embedding maps with time evolution maps,  a path independence requirement of the time evolution maps.

Given such a system, we can define the continuum limit of the theory, as an  inductive limit. 
This limit is defined as the space of equivalence classes  ${\cal H}:=\cup_b {\cal H}_b \slash \sim$.  The equivalence relation is defined as follows: two states $\psi_b$ and $\psi'_{b'}$ are equivalent, if there exist a  $b''$ with $b\prec b''$ and $b'\prec b''$, i.e. a discretization $b''$ refining both $b$ and $b'$, such that $\iota_{bb''}(\psi_b)=\iota_{b'b''}(\psi_b')$.

 In words, two states on different discretizations $b,b'$ are equivalent, if they can be refined to the same state. This notion of inductive limit allows to embed any `discrete state' $\psi_b$ into the continuum Hilbert space ${\cal H}$ via an embedding $\iota_b$.

As mentioned this construction is used in loop quantum gravity on the kinematical level, that is the choice of embedding maps is not tied to a dynamics.
 Indeed, in a theory with a proper time evolution one would need to separate the refining time evolution steps into a `purely refining' part and a `proper evolution' part, as in the example in section \ref{scalar}. Otherwise one would identify time evolved states as equivalent.

However, in gravitational theories, time evolution is a gauge transformation. In the quantum theory, the time evolution operator (\ref{pathintegral}) is supposed to act as a projector onto physical states and thus as an identity on physical states. 
Hence one can indeed attempt to use refining time evolution, to define the embedding maps $\iota_{b b'}$ between different discretizations. As we will discuss, the difficulty is that refining time evolution maps based on `naive' discretizations, will not satisfy the consistency conditions (\ref{cons1}). Here coarse graining provides a means to reach theories in which the consistency conditions are actually satisfied.

\subsection{Refining time evolution and no--boundary vacuum} \label{timeevol}

Let us return to the time evolution operator (kernel) defined from the path integral (\ref{pathintegral}) 
\ba\label{pathintegral2}
K(X_{ini},X_{fin})&=& \int_{X_{ini},X_{fin} \text{fixed} } {\cal D} X \, \exp\left( \frac{i}{\hbar} S(X)\right) \q .
\ea
Here we denote by $X_{ini}$ and $X_{fin}$ initial and final configuration data. In, for instance simplicial, discretizations of the path integral (\ref{pathintegral2})  the wave functions $\psi_{i}(X_{i})$ and $\psi_{f}(X_{fin})$  might be from two Hilbert spaces ${\cal H}_{b_i}$ and ${\cal H}_{b_f}$ associated to two different discretizations $b_i$ and $b_f$.

Even if we consider a system with proper time evolution the path integral (\ref{pathintegral2}) will project onto states satisfying the pre-- and post constraints discussed in section \ref{class}. The reason is similar to the mechanism turning path integrals for gauge theories into projectors \cite{proj}, we will sketch an argument here, valid for linearized theories \cite{tedtoappear}:  Consider for instance the case of post--constraints $C_i(X,P)$, where $P$ are the momentum variables conjugated to $X$. 

We discussed in section \ref{class} that these post--constraints are first class and lead to post--gauge degrees of freedom, that is part of the configuration data $X$ at final time remain undetermined.  On the other hand there will be also post--Dirac observables, i.e. functions on phase space that Poisson commute with the post--constraints $C(X,P)$. The structure of the constraints allows to make a canonical variable transformation such that the configuration variables separate into post gauge $X^G$ and post--Dirac $X^D$ degrees of freedom. The constraints then involve only variables conjugated to the post--gauge variables $X^G$ and the variables $X^G$ themselves.

By integrating over all bulk variables in (\ref{pathintegral2}) we can define an effective action that only depends on initial and final configuration variables: 
\ba\label{pathintegral3}
K(X_{ini},X_{fin})&=& \exp\left( \frac{i}{\hbar} S_{eff}(X_{ini}, X_{fin})\right)
\ea
We can use the canonical variable transformation for the final configuration data $X_{fin}$. The fact that the classical action leads to post--constraints means that the effective action decouples  gauge and Dirac degrees of freedom 
\ba
S_{eff}\,=\,S^D(X_{ini},X^D_{fin}) \,+\, S^G(X^G_{fin}) \q.
\ea
This makes the appearance of constraints $C(X^G_{fin}, P^G_{fin})$ obvious
\ba
P_{fin}^G\,=\,-\frac{\partial S_{eff}}{\partial X^G_{fin}} \,=\, -\frac{\partial S^G(X^G_{fin})}{\partial X^G_{fin}}  \q .
\ea
The time evolution kernel (\ref{pathintegral3}) is therefore of the form
\ba
K(X_{ini},X_{fin})&=& \exp\left( \frac{i}{\hbar} S^G( X^G_{fin})\right) \,\times \,  \exp\left( \frac{i}{\hbar} S^D( X_{ini},X^D_{fin})\right)  \q .
\ea
All states resulting from a time evolution
\ba\label{result}
\psi_f(X_{fin})&=& \int \bd X_{ini} \,  \exp\left( \frac{i}{\hbar} S^G( X^G_{fin})\right) \,\times \,  \exp\left( \frac{i}{\hbar} S^D( X_{ini},X^D_{fin})\right) \,\psi_{i}(X_{ini})
\ea
have a prescribed factor $ \exp\left( \frac{i}{\hbar} S^G( X^G_{fin})\right)$ determining the dependence of the wave function in the gauge variables. Adopting a  Schroedinger quantization scheme, with the momenta quantized as derivative operators $\hat P=  \partial/\partial X$ and configurations as multiplication operators, the states (\ref{result}) satisfy the quantized constraints
\ba
\hat C &=& -i\hbar \frac{\partial }{\partial X^G_{fin}}  + \frac{\partial S^G(X^G_{fin})}{\partial X^G_{fin}} \q .
\ea

In summary the choice of discrete action for a refining time evolution leads to constraints that determine the behaviour of the resulting wave functions in the `finer' degrees of freedom, which here are characterized as post-gauge degrees of freedom. 

As mentioned this mechanism holds also for theories which a priori do not show any gauge symmetries and thus we deal with a proper time evolution operator. General relativity is a totally constraint system. Formal arguments show that the path integral (\ref{pathintegral2}) is equivalent to a projector onto the Hamiltonian and diffeomorphism constraints $C_I$ of the theory \cite{proj}
\ba\label{timeev}
\int  {\cal D} N^I  \exp\left( \frac{i}{\hbar} N^I \hat C_I \right)   \q .
\ea
Here $N^I$ denote Lagrange multiplier, known as lapse and shift. The integration over these multipliers induces a averaging over the action of the Hamiltonian and diffeomorphism constraints. 
For a discussion of  the many subtleties involving this proposal see for instance  \cite{alex3d,thomasbook,master}.)  The averaging would therefore project onto states that satisfy the constraints.

In our discrete context, allowing for the possibility of discretizations changing in time, one expects that the Hamiltonian and diffeomorphism constraints will be part of the post-- or pre-- constraints. As mentioned this issue is however involved, as discretizations typically break diffeomorphism symmetry, which leads to the constraints. For the moment we will ignore this issue and comment later how to deal with it.

Thus we can hope\footnote{In fact, a naive discretization will break diffeomorphism symmetry and thus the statement regarding (a) can hold only in some approximate sense.} that a simplicial discretization of a path integral describing refining time evolution will lead to states which (a) satisfy the Hamiltonian and diffeomorphism constraints and (b) in which the finer (Dirac) degrees of freedom are also put into a specific state, characterized by the remaining post-- constraints.

With a simplicial path integral, we can in particular consider the extreme case of a refining time evolution; that is, we can start with zero-dimensional configuration space and evolve to a large triangulated spherical hypersurface \cite{hoehn2}. That is the first evolution step evolves from a vertex to the boundary of a $d$--dimensional simplex, where $d$ denotes the space time dimension. TThe wave function will be given only as the (path integral) amplitude associated to this simplex. The following evolution steps can be understood as gluing further simplices to the one we started with, by multiplying the wave function with the corresponding simplex amplitudes and integrating over all variables that become bulk. 

In this case we will have at every step as many post constraints as (configuration) variables, i.e. the reduced phase space is zero--dimensional. 
 Indeed all momenta $P_b$ are generated by Hamilton's principal function $S_H$, i.e. the action evaluated on a solution prescribed by the boundary configurations $X$:
\ba\label{mom}
P &=& \frac{\partial S_H}{\partial X}(X) \q .
\ea
We thus  have constraints $C=P-\frac{\partial S_H}{\partial X}$.  These are Abelian, as the momenta are coming from a generating function. The phase space is foliated by gauge orbits, generated by the constraints, i.e. all configurations $X$ are post--gauge degrees of freedom.

 In the quantum theory this corresponds to a unique physical wave function\footnote{In fact, this wave function will in general depend on the underlying discretization, which can be interpreted as a choice of order for refining time evolution maps. Thus proper `uniqueness' requires a notion of path independence, as will be explained in section \ref{pathdep}}, given by a (Hartle Hawking) no--boundary vacuum \cite{hhv}. In the semi--classical approximation we have
\ba\label{HHv}
\psi_{HH}(X) \sim  \exp \left(\frac{i}{\hbar} S_H \right) \q .
\ea
%The resulting state can be interpreted as a representation of the  vacuum state on the given triangulation. We will elaborate on this notion in section \ref{dynemb}.

Here one would indeed expect the appearance of the standard vacuum, at least in the limit of infinitely large regions \cite{florian}. Gravitational theories play a special role here, as the size of the region is encoded in the state itself, thus the wave function gives rather a probability distribution for the geometrical volume of the hypersurface.   `Radial' evolution, as described here should not change physical states as it is just another form of time evolution.  Thus we can hope that a  vacuum is reached for degrees of freedom describing scales (much) larger than the discretization scale of the boundary.

We note that a framework, which permits time evolution with phase spaces or Hilbert spaces that change in time, allows to define a notion  of  vacuum. For instance starting with a very coarse state and refining this state in an homogeneous manner should result into a state describing homogeneous geometries. This allows for applications for cosmology based on lattice treatments, for instance \cite{bojowald}.

An interesting question for future research will be to investigate which simplicial quantum gravity models will lead to an acceptable (Hartle Hawking) vacuum and to investigate the properties of this vacuum.  

Apart from defining a no--boundary wave function, the refining time evolution can of course also be used to refine states -- and thus to provide the embedding maps needed for the construction of inductive limit Hilbert spaces, as discussed in section \ref{emb}.  
Such (dynamical)  embedding maps are therefore selected by taking the dynamics of the theory into account, which is particularly advised for coarse graining \cite{bd12b}. 
Here one has however to address the issue that discretized path integrals will in general break diffeomorphism symmetry and, related to this fact, be triangulation dependent.  This will be subject of the next sections.

\subsection{Path independence of evolution and consistent embedding maps}\label{pathdep}

We argued that a discrete evolution starting from a zero--dimensional phase space or a one dimensional Hilbert space produces a vacuum state. 
However this vacuum state will in general depend on the order of the time evolution steps, which for a simplicial  discretization determines the triangulation of the bulk that is bounded by the triangulated hypersurface on which the vacuum is defined.

Similarly, if we aim to use the refining time evolution defined by the path integral as embedding maps, the consistency conditions \eqref{cons1} will in general not be satisfied. These consistency conditions can now be interpreted as demanding independence of the evolved state from the chosen evolution path. It  can be understood as a discrete version of implementing the Dirac algebra of (Hamiltonian and spatial diffeomorphism) constraints. As pointed out in \cite{kuchar} the Dirac algebra implies path independence, with respect to evolving through arbitrary choices of spatial hypersurfaces. This constitutes a further\footnote{
In cases where diffeomorphism symmetry is realized, for instance in 3D discrete gravity, 4D gravity restricted to the `flat' sector \cite{ryan}, or 4D linearized gravity \cite{hoehn1}, one can also introduce a continuum time evolution generated by Hamiltonian constraints \cite{commi,bd08,bahrdittrich09a} and define a (first class) Dirac algebra of these constraints \cite{bonzomDirac}. This continuum time evolution reproduces the discrete time evolution \cite{bd08,hoehn2}, if one integrates the infinitesimal evolution to one with a finite time.} relation between  diffeomorphism symmetry, that yields the constraints,  and triangulation independence \cite{steinhaus11}.

 So far we discussed only consistency for the embedding maps, which is needed to make the projective limit Hilbert space well defined.
 Observables on this Hilbert space need also to satisfy  conditions  known as  cylindrically consistency: Observables ${\cal O}_b$ defined on the family of Hilbert spaces ${\cal H}_b$ need to commute with the embedding maps $\iota_{bb'}$:
\ba\label{obs}
\iota_{bb'} ({\cal O}_b\, \psi_b) \,=\, {\cal O}_{b'}  \, \iota_{bb'} (\psi_{b})
\ea
for all states $\psi_b \in {\cal H}_b$ and all pairs $b\prec b'$. This ensures that the observables are well defined on the continuum Hilbert space, i.e. do not depend on the representative  $\psi_b$ chosen.
In the case that $\iota_{bb'}$ is given by a refining time evolution consistent observables have therefore to be `refining Dirac observables'.  The algebra of refining Dirac observables characterizes the resulting continuum Hilbert space, as it provides a representation of this algebra.

 Topological theories  can often be discretized such that partition functions and physical observables are triangulation independent. This also includes 3D gravity, which is topological. Due to the triangulation invariance  the refining time evolution maps defined via the discretized path integral do satisfy the consistency conditions (\ref{cons1}).  We will illustrate this situation in section \ref{top}.  As we will comment there, the set of `refining Dirac observables' is much bigger than the set of (standard) Dirac observables given by the topological theory. This allows to use refining time evolution maps stemming from topological theories to define (kinematical) Hilbert spaces for other theories. They can also be used to construct a new Hilbert space for loop
quantum gravity, based on the time evolution map of $BF$--theory \cite{geiller}.

We believe that the application of refining time evolution maps is however not restricted to topological theories, despite the challenges posed by the triangulation dependence of the path integral. The strategy to attack this issue is to improve a given discretization by coarse graining. The fixed point of the coarse graining flow is hoped to show enhanced symmetry properties, in particular diffeomorphism symmetry which is tied to triangulation dependence \cite{steinhaus11,bd12a}.  

Such a coarse graining flow leads however to non--local couplings\footnote{Triangulation invariant theories with local couplings are always topological theories, see for instance \cite{hoehn3,bdwk}.}, which are difficult to control. One would then also  expect the embedding maps, if defined via refining time evolution, to be highly non--local. In  section \ref{howto} we will discuss a coarse graining framework which avoids this issue, and moreover is based on the concepts introduced so far.

Let us comment on the appearance of discretization changing time evolution in loop quantum gravity. There   graph changing (actually graph refining) Hamiltonian constraints have been defined by Thiemann  \cite{thiemann}. These constraints are anomaly free, in the sense that the commutator of two Hamiltonians vanishes if  evaluated on the Hilbert space of diffeomorphism invariant states, see \cite{thiemann,habitat,madha} for discussions. 
 
What is missing is a concrete geometric interpretation of the action of these constraints and a concrete connection to the path integral. (The notion of graph changing Hamiltonians inspired the development of spin foams, as time evolved spin networks  \cite{reisenbergerrovelli}.) 
This discussion here suggest a possible interpretation for the graph changing Hamiltonians, the exponentiation of which should lead to a (refining) time evolution.  Thus one could attempt to extract a notion of vacuum from the Hamiltonian constraints. 

%In section \ref{scalar} we provided a classical model that gives an intuition of graph changing time evolution, and might help to understand graph changing Hamiltonians in more general situations.

%One might wonder how  graph changing constraints relate to a graph changing time evolution. The notion of dynamical cylindrical consistency suggests that the states generated by the constraints, and therefore based on different discretizations, should be in the same equivalence class, i.e.  identified with each other.

\subsection{Pre--constraints and coarse graining}

We suggested to use the refining time evolution to define embedding maps for the inductive Hilbert space construction. Refining time evolution leads to post--constraints, which we argued characterize the (vacuum) state, into which the finer degrees of freedom are put. Here we want to comment shortly on the role of pre--constraints.

These appear for coarse graining time evolution steps, that is the number of variables decreases. 
Classically these constraints demand that a state needs to satisfy certain conditions, so that the time evolution move can be applied. By time inversion symmetry we can understand this condition in the following way: the state has to be equivalent to a refining of a coarser state. Although the state is represented on a fine triangulation it does only include  degrees of freedom in non--vacuum states on a coarser scale.  

Although for classical evolution one has to satisfy these  constraints,  quantum mechanical evolution is always possible. This also holds for standard gauge systems: a priori a quantum state does not need to satisfy any constraints to serve as a boundary condition in a path integral. Rather the path integral itself will project out non--physical degrees of freedom \cite{proj}. Thus, one can indeed expect that in a quantum evolution, the degrees of freedom which are too fine to be evolved classically (as identified by the constraints) will be projected to the vacuum.  In this sense the quantum mechanical evolution is automatically providing a coarse graining. 
Note that this will be a non--unitary evolution as it includes a projective part. A unitary description can only be obtained if one restricts to the subspace of the Hilbert space describing only sufficiently coarse degrees of freedom, i.e. the physical Hilbert space with respect to the pre--constraints, see also \cite{hoehnq}.

Thus time evolution cannot be inverted: Concatenating coarse graining and refining  we isolate the projective part, that can be understood as projecting fine degrees of freedom to the vacuum state, that is the state obtained will automatically satisfy the initial pre--constraints. 
This provides an interesting asymmetry in time evolution that might serve as an arrow of time. See \cite{shape} for another proposal on the origin of the arrow of time, in which also the notion of complexity of the state is crucial.

\section{How to define a continuum theory of quantum gravity}\label{howto}

The investigation of time evolution with changing phase space or Hilbert space dimension is motivated by the simplicial discretization of gravity \cite{hoehn1,hoehn2,hoehn3}. However, such discretizations break diffeomorphism symmetry for the 4D theory  \cite{bd08}. This appears both at the classical level \cite{bahrdittrich09a} , and even in more severe form on the quantum level. For instance, the classical 4D Regge action, is invariant under 5--1 moves, but not under 3-3 moves. The latter fact can be related to a breaking of diffeomorphism symmetry on the classical level. Moreover on the quantum level one can show that no local path integral measure factor exists that makes the theory invariant under 5--1 moves \cite{sebmeas,nonlocal}, implying even a breaking of the residual classical symmetry.

This implies in particular that the consistency conditions formulated in (\ref{cons1}) are violated. A way out is to improve the discretization by coarse graining, see  \cite{bahrdittrich09b,steinhaus11} for examples. At fixed points of the coarse graining flow one might arrive at perfect discretizations \cite{hasenfratz}, for which consistency conditions of the form (\ref{cons1}) are satisfied.  These fixed points represent the continuum limit of the theory one started with, however expressed on a discretization. 

There are different ways to proceed with the coarse graining.  One is to keep basic building blocks but to allow highly non--local couplings, which are naturally induced by the coarse graining \cite{bietenholz,he}. As was pointed out in \cite{bd12b}, there is an alternative inspired by tensor network renormalization (which we will explain in section \ref{tensor}) and the generalized boundary proposal \cite{oeckl}.

This alternative construction of a consistent theory would not put basic building blocks (with simplest possible boundary discretizations) with their amplitudes in the centre but instead amplitude maps for space time regions, with arbitrarily complicated discretization of the boundary.  These amplitude maps are built from the basic amplitudes, and agree basically with the (dual of the) Hartle Hawking no boundary wave function.  The amplitude maps are defined on Hilbert spaces ${\cal H}_b$ associated to the discretized boundaries $b$ of  a space time region: ${\cal A}_b:{\cal H}_b \rightarrow {\mathbb C}$ as
\ba\label{defam}
{\cal A}_b(\psi_b)&:=&  \int {\cal D} X {\cal D} X_b \exp \left(\frac{i}{\hbar} S (X,X_b) \right) \,\psi_b (X_b) \nn\\
&=& \langle \psi_\emptyset  | ({\mathbf K}_{\emptyset b})^\dagger | \psi_b \rangle \,= :\, \langle \psi_\emptyset  | \psi_b \rangle_{phys}
\ea
where we denote the bulk configuration variables  with $X$ and the boundary variables with $X_b$. Thus the amplitude map applied to the wave function $\psi_b$ is given by the inner product between this wave function $\psi_b$ and the no--boundary wave function. This no--boundary wave function is here expressed as time evolution operator ${\mathbf K}_{\emptyset b}$ applied to the one--dimensional wave function $\psi_\emptyset$ associated to the empty discretization. The second line in (\ref{defam}) defines the physical inner product, between the projections onto physical states of the two (kinematical) states $\psi_\emptyset$ and $\psi_b$.

As usual the path integral in (\ref{defam}) is a discretized one. Thus the first task is to arrive at amplitude functionals ${\cal A}_b$  for fixed boundaries $b$ that are independent of the bulk triangulation. One way to reach such amplitudes is by coarse graining, as will be explained in the next section \ref{tensor}. 

For  a very coarse boundary $b$ we can triangulate the bulk with very few simplices. For instance the boundary of a simplex can be triangulated with just this simplex and thus the amplitude functional ${\cal A}_b$, discretized in this way, would be just given by the pairing of the simplex amplitude with the boundary wave function. However there are infinitely many ways to subdivide this simplex (keeping the boundary), and thus one would have to find a method to determine the actual ${\cal A}_b$. Indeed we will specify a further criterion for these amplitudes, which will actually help to construct the coarse graining flow of these amplitudes.

It is important to note that bulk triangulation independence of the amplitude maps ${\cal A}_b$ is {\it not} sufficient for the construction of the continuum limit. (Indeed one could just declare some rule for selecting a particular bulk triangulation for each boundary.) We rather need to demand  a condition that connects the amplitude maps ${\cal A}_b$ for  different boundaries $b$.

Thus we need first to choose embedding maps $\iota_{bb'}$ that connect the different boundary Hilbert spaces, as explained in section \ref{emb}. As we explained, there might be different sets of embedding maps, leading to different continuum Hilbert spaces. We will see that some choices are preferred over others. With a given choice of embedding map we require 
 that the amplitude maps are cylindrically consistent functionals, that is 
\ba
{\cal A}_{b'}(\iota_{bb'}(\psi_b)) &=& {\cal A}_b(\psi_b) \q .
\ea
In words, if we take a coarse state and evaluate the corresponding amplitude map ${\cal A}_b$ on it, we should get the same result as first embedding the state into the `finer' Hilbert space ${\cal H}_b$ and then evaluating with the `finer' amplitude map ${\cal A}_{b'}$. Thus, the result should not depend on which boundary we choose to represent the equivalence class of states $[\psi_b]$ under the equivalence relations of the inductive limit. This allows to actually define the amplitude map as a functional on the inductive  (i.e. continuum) limit Hilbert space ${\cal H}$ defined in section \ref{emb}.  Such a requirement was proposed in \cite{bahrproc} with regard to the (kinematical) embedding maps of the Ashtekar Lewandowski Hilbert space. We will argue here that the construction of cylindrically consistent amplitudes is facilitated by the adoption of dynamical embedding maps, as provided by refining time evolution.

The amplitude map ${\cal A}_{[b]}$  is technically not any more labelled by a discretization as such, but by equivalence classes of discretizations. Here two discretizations are equivalent if they can be refined to the same discretization. Thus, the information that is left over could just carry topological information (for gravitational theories where metric variables are dynamical) of the boundary. 

In our case we assumed spherical topology, thus a cylindrical consistent family of amplitude maps defines a continuum amplitude ${\cal A}$. This amplitude ${\cal A}$ replaces the basic amplitude for, say the boundary of a simplex, one starts with in the regularization of the path integral. We can recover a `perfect' amplitude, by evaluating ${\cal A}$ on states that are equivalent to states defined on a simplex boundary under the chosen embedding map. 

The cylindrically consistency requirement for the amplitude maps is a very strong requirement -- it basically encodes the solution of the theory. We can hope to build such amplitude maps iteratively, for more and more refined boundaries, as will be the subject of the next section. To this end it is important to choose embedding maps that are adapted to the dynamics of the system \cite{bd12b}. In particular we suggested that refining time evolution should give good embedding maps. A priori these will typically fail to satisfy the consistency requirement (\ref{cons1}). However the improved amplitude maps ${\cal A}_b$ also allow to define an improved discretization of the path integral and thus to define a (refining) time evolution, that will satisfy the consistency requirement to a better approximations\footnote{The consistency equations can be tested if one considers the equations on matrix elements $\langle \psi_{b''} |\iota_{bb''} (\psi_b)\rangle \stackrel{!}{=}\langle \psi_{b''} |\iota_{b'b''}\circ \iota_{bb'} (\psi_b)\rangle$. Under an iterative improvement the equations will be satisfied  for a larger and larger class of states involving finer and finer boundaries.}. In an iterative process one therefore improves both the amplitude maps and the embeddings, if these are defined by refining time evolution.

The reason why such embeddings are particularly apt to define cylindrically consistent amplitudes is in the definition of the amplitudes in (\ref{defam}). If $\iota_{bb'}={\mathbb K}_{bb'}$ we will have
\ba\label{iter1}
{\cal A}_{b'}(\iota_{bb'} \psi_b)&=& \langle \psi_\emptyset  | ({\mathbf K}_{\emptyset b'})^\dagger  | {\mathbf K}_{bb'} \psi_b \rangle \, \sim\, 
\langle \psi_\emptyset  | ({\mathbf K}_{\emptyset b})^\dagger   | \psi_b \rangle 
\,=\, {\cal A}_{b}(\psi_b) \q .
\ea
Here we wrote $\sim$ in the second equation as $({\mathbf K}_{\emptyset b'})^\dagger \circ {\mathbf K}_{bb'}  \sim ({\mathbf K}_{\emptyset b})^\dagger$ holds only approximately in the discretization. However we see that embedding maps defined via refining time evolution simplify the task of constructing cylindrically consistent amplitudes. Indeed the consistency condition for these embedding maps are tied to the cylindrical consistency of the amplitudes.

We want to remark 
 that we do {\it not} require a consistent gluing between the cylindrically consistent amplitude maps as long as this gluing is performed on a discrete boundary $b$. (That is the gluing involves integration only over the variables $X_b$.) 
 One could for instance require that the amplitude for a region with a more complicated boundary ${\cal A}_{b_3}$ arises as the gluing between two amplitudes with less refined boundaries ${\cal A}_{b_1}$ and ${\cal A}_{b_2}$, similar to the way one would glue simplices together. We expect that such a relation might indeed hold, however only if one preforms a continuum limit for the piece of boundary that is glued over.

Let us emphasize that the amplitude maps ${\cal A}_{[b]}$ are the end point of a construction to reach the continuum limit of the theory. Of course one hopes that the `initial' theory defined via basic building blocks and local couplings, provide the basis for the construction of such a theory. This implies that this `initial' theory can nevertheless be used to extract sufficiently coarse grained observables from sufficiently refined discretizations. 

As mentioned we aim at constructing both, cylindrically consistent amplitudes and consistent embedding maps given by refining time evolution. In the next section \ref{tensor} we will explain that tensor network coarse graining tools provide methods to construct these.

%__________________________

\section{Tensor network coarse graining: time evolution in radial direction}\label{tensor}

Tensor network renormalization group methods \cite{no,levin,guwen,others} can be understood to implement an iterative method to construct cylindrically consistent amplitudes. Coming back to equation (\ref{iter1})
\ba\label{iter2}
{\cal A}_{b'}(\iota_{bb'} \psi_b)&=& \langle \psi_\emptyset  | ({\mathbf K}_{\emptyset b'})^\dagger  | {\mathbf K}_{bb'} \psi_b \rangle \, \sim\, 
\langle \psi_\emptyset  | ({\mathbf K}_{\emptyset b})^\dagger   | \psi_b \rangle 
\,=\, {\cal A}_{b}(\psi_b) \q .
\ea
we can understand the second term to consist of two parts: the first is the computation of  $\langle \psi_\emptyset  | ({\mathbf K}_{\emptyset b'})^\dagger$, that is the basically the amplitude functional ${\cal A}_{b'}$ for a more refined boundary. One would build such an amplitude functional from gluing amplitudes ${\cal A}_b$ for less refined boundaries $b$. 

However we want to define an iterative process that improves the amplitude maps ${\cal A}_b$, which are functionals on ${\cal H}_b$. We thus have to find a way to  pull back the amplitudes ${\cal A}_{b'}$ to ${\cal H}_b$, which is done by using the embedding map $\iota_{bb'}={\mathbf K}_{bb'}$. Thus one defines the improved amplitudes ${\cal A}^{imp}_b$ as
\ba
{\cal A}_{b}^{imp} (\psi_b) &=&  \langle \psi_\emptyset  | ({\mathbf K}_{\emptyset b'})^\dagger  | {\mathbf K}_{bb'} \psi_b \rangle \q .
\ea
Here both $({\mathbf K}_{\emptyset b'})^\dagger $ and ${\mathbf K}_{bb'}$ are built from using the initial ${\cal A}_b$ as basic amplitudes. 

The process is repeated for the improved amplitudes ${\cal A}_{b}^{imp}$ until the procedure converges to a fixed point ${\cal A}_{b}^{fix}$. This fixed point amplitude can be used to proceed to a more refined pair of boundaries $(b',b'')$ with $b\prec b''$ to find the next fixed point amplitude ${\cal A}^{fix}_{b'}$ and so on.

There are many  tensor network renormalization algorithms \cite{no,levin,guwen,others}, which differ in their geometric setup and the details of how to define $({\mathbf K}_{\emptyset b'})^\dagger$ and the embedding $\iota_{bb'}$.  We will shortly explain a method that can be interpreted as radial evolution, as this also matches nicely with amplitudes being defined via the no--boundary wave function, as in (\ref{defam}).

The name tensor networks refers to the fact that the amplitudes of a space time region are encoded in tensor of a given rank $n$, associated to an $n$--valent vertex, which we can imagine to sit inside this space time region. The indices of this tensor encode the boundary data of the space time region, hence contracting tensors of two neighbouring regions corresponds to gluing the associated amplitudes.  The rank $n$ and the bond dimension $\chi$ (equal to the number of values the index can take) determines the amount of boundary data and hence the fineness of the boundary in question. Note that one can redefine higher rank tensors to tensors of lower rank by summarizing for instance two indices $(i,j)$  with $\chi_i,\chi_j$ into effective indices $I=(i,j)$ with bond dimension $\chi_I=\chi_i \cdot \chi_j$.

This interpretation matches nicely with spin nets \cite{sffinite} and spin foams describing gravitational dynamics. The former can be naturally understood as tensor networks  \cite{eckert,bdetal13a,bdetal13b}. A tensor network description of spin foams can be found in \cite{eckert}.

\subsection{Radial evolution}

As we will see tensor network algorithms are related to transfer matrix methods in which the (Wick rotated) time evolution operator is diagonalized. For the latter, Wick rotation is essential, as the eigenvalues of the transfer operator need to be ordered in size; in this way we can distinguish relevant from irrelevant degrees of freedom. However one can understand tensor networks to replace the time evolution operator with a radial evolution operator. Even if the (standard) time evolution operator might be unitary, and hence all eigenvalues with absolute value equal to one, the radial evolution operator will include a projective part, that -- as we have argued will project out finer degrees of freedom. This can then be used for the construction of an embedding map.

An evolution in radial direction is also expected to project onto the vacuum state \cite{florian,levin}, see also the discussion in section \ref{timeevol} which involves the non--Wick rotated amplitudes.\footnote{For a (Wick rotated) time evolution operator $\exp(-\int_0^R H_r \bd r )$ acts as a projector on the ground states of the  Hamiltonian $H$  for $R$ going to infinity. Here $H_r$ denotes the Hamiltonian for radial evolution at the radius $r$. 
% which is built as a sum over a basic two site Hamiltonian $h_i$ (that is we have nearest neighbour coupling). 
For large radius, we will have small $dr/r$, and hence $H_r$ approaches the Hamiltonian $H$ for the time evolution of constant volume hypersurfaces. }

\begin{figure}
\begin{center}
\begin{tikzpicture}[scale=0.8]
\draw (-0.5,0) -- (0,0) -- (0.5,0)
      (0,-0.5) -- (0,0) -- (0,0.5);
\draw (0,0) node {\textbullet};
%\draw [dashed] (0,0) circle [radius =0.3];
\draw [dashed] (0.3,0) arc(0:-180:0.3) arc(180:0:0.3);
\draw (3.5,0) -- (4,0) -- (4.5,0) -- (5,0) -- (5.5,0) -- (6,0.) -- (6.5,0.)
      (3.5,-1) -- (4,-1) -- (4.5,-1) -- (5,-1) -- (5.5,-1) -- (6,-1) -- (6.5,-1)
      (3.5,1) -- (4,1) -- (4.5,1) -- (5,1) -- (5.5,1) -- (6,1) -- (6.5,1) 
      (5,-1.5) -- (5,-1) -- (5,-0.5) -- (5,0) -- (5,0.5) -- (5,1) -- (5,1.5)
      (4,-1.5) -- (4,-1) -- (4,-0.5) -- (4,0) -- (4,0.5) -- (4,1) -- (4,1.5)
      (6,-1.5) -- (6,-1) -- (6,-0.5) -- (6,0) -- (6,0.5) -- (6,1) -- (6,1.5);
\draw (5,0) node {\textbullet};
\draw [gray] (4,0) node {\textbullet}
             (6,0) node {\textbullet}
             (4,1) node {\textbullet}
             (4,-1) node {\textbullet}
             (5,1) node {\textbullet}
             (5,-1) node {\textbullet}
             (6,1) node {\textbullet}
             (6,-1) node {\textbullet};
% \draw [dashed] (5,0) circle [radius =1.6];
\draw [dashed] (6.5,0) arc(0:-180:1.5) arc(180:0:1.5);
\draw[->] [thick] (1,0) -- (3,0);
\draw[->] [thick] (4.7,0.3) -- (4.2,0.8);
\draw[->] [thick] (4.7,-0.3) -- (4.2,-0.8);
\draw[->] [thick] (5.3,0.3) -- (5.8,0.8);
\draw[->] [thick] (5.3,-0.3) -- (5.8,-0.8);
\end{tikzpicture}
\caption{Illustration of radial time evolution in tensor networks: By adding eight additional tensors (in gray) we perform one time evolution step. The boundary data grows exponentially from $\chi^4$ to $\chi^{48}$. \label{fig:radial}}
\end{center}
\end{figure}
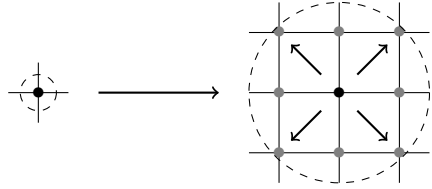

Consider a radial evolution as in figure  \ref{fig:radial}. Here the amplitude / tensor for a larger region is built from the amplitude/ tensor of a basic building block, represented by a (dual) vertex.
One would now like to treat the amplitude for the new region as an effective tensor  and repeat the procedure. However one has to face the problem, that the number of boundary data, grows exponentially during this procedure, making it impossible to implement in practise. 

We thus have to find a method to project back the amplitudes to a boundary with less data, i.e. to coarser boundaries.
The radial time evolution can be split into steps  with time evolution operators 
\be
T(R_1,R_2)=\exp(-\int_{R_1}^{R_2} H_r \bd r ) \q .
\ee
This is a refining time evolution in the sense that the Hilbert space  ${\cal H}_2$ at $R_2$ will have more kinematical degrees of freedom than the Hilbert space ${\cal H}_1$ at $R_1$. Thus $T(R_1,R_2)$ will have a projective part (even if we would not have Wick rotated), which can be identified by a singular value decomposition. This would give a maximal number of $\text{dim} ({\cal H}_1)$ singular values. Hence  $T(R_1,R_2)$ will have a non--trivial co--image in ${\cal H}_2$, which can be projected out. A new amplitude can therefore be defined on a $\text{dim} ({\cal H}_1)$ subspace, which can be identified as the subspace carrying coarse boundary data.

Such a scheme might be indeed worthwhile to investigate further (for statistical systems), in order to obtain an intuition about the truncations.  One would however expect that the reorganization of the degrees of freedom via the singular value decomposition will be highly non--local, as we have seen for the example in section \ref{subscalar}. The time evolution considered there exactly corresponds to $T(R_1,R_2)$.  This non--locality makes it however difficult to turn this into an iterative procedure. The new amplitudes will be expressed with respect to data spread over the entire boundary, which makes a local gluing of these amplitudes difficult.

%As a remark, we now see why tensor network coarse graining should also work for theories in which evolution between equal size Hilbert space is actually unitary. (For statistical systems one usually works with $\exp(-H)$, so the path integral is Wick rotated. Here we refer to a non--Wick rotated path integral.) In such a case one might argue that all singular values of the transfer matrix should be equal to one. However adopting the picture of radial evolution, we do not have a unitary evolution -- one rather expects an isometric embedding of coarser states into a Hilbert space of finer states. Thus the transfer matrix should lead to a set of singular values equal to one and others vanishing. A good truncation would then identify these vanishing singular values. In this way the singular value decomposition and the associated  embeddings $V$ can rather be understood as a field redefinition, which automatically defines appropriate coarse grained observables of the theory.

\subsection{Truncations via singular value decomposition} 

\begin{figure}
\begin{center}
\begin{tikzpicture}[scale=0.55]
\draw [thick] (0,0) rectangle (3,3)
      (5,0) rectangle (8,3);
\draw (1.5,1.5) node {{\huge $M$}}
      (6.5,1.5) node {{\huge $M$}};
\draw (3,1) -- (5,1)
      (3,2) -- (5,2)
      (0,2.5) -- (-1,2.5)
      (0,2) -- (-1,2)
      (-0.5,1.5) node {{\large $\vdots$}}
      (0,1) -- (-1,1)
      (-2,1.5) node {\Large $A$}
      (8,2.5) -- (9,2.5)
      (8,2) -- (9,2)
      (8.5,1.5) node {{\large $\vdots$}}
      (8,1) -- (9,1)
      (10,1.5) node {\Large $B$}
      (4,0.6) node {{\Large $\beta$}}
      (4,2.25) node {{\Large $\alpha$}};
\end{tikzpicture}  \quad \quad 
\begin{tikzpicture}[scale=0.5]
\draw [thick] (0,0) rectangle (3,3);
\draw (1.5,1.5) node {{\huge $M$}};
\draw (3,1) -- (5,1) arc(-90:90:0.5)
      (3,2) -- (5,2)
      (5.5,1.5) -- (7,1.5)
      (5.5,1.5) node {\large \textbullet}
      (7,1.75) node {\Large $i$}
      (5.9,0.9) node {\Large $V$}
      (0,2.5) -- (-1,2.5)
      (0,2) -- (-1,2)
      (-0.5,1.5) node {{\large $\vdots$}}
      (0,1) -- (-1,1)
      (-2,1.5) node {\Large $A$}
      (4,0.6) node {{\Large $\beta$}}
      (4,2.25) node {{\Large $\alpha$}};
\end{tikzpicture}
\caption{Left: Two regions in a tensor network, encoded in the matrices $M$, are sharing two edges with labels $\{\alpha,\beta\}$, which have a total range of $\chi^2$. Right: From the singular value decomposition we can define the map $V$ depicted as a three--valent vertex, where we restrict the label $i$ of the singular values to be $\leq \chi$. \label{fig:matrix-svd}}
\end{center}
\end{figure}
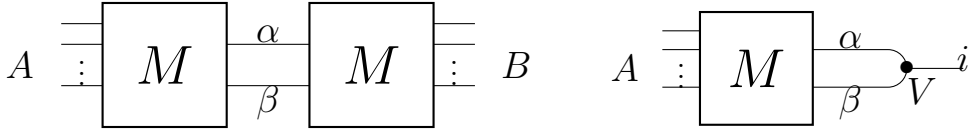

In practice one therefore employs schemes which involve more local truncations.
The basic idea is as follows. Imagine two space time regions or effective vertices connected with each other by two edges, representing the summation over a certain set of variables, see figure \ref{fig:matrix-svd}. We would like to replace these edges carrying an index pair $\{\alpha,\beta\}$ of size $\chi^2$ with an effective edge carrying only a number $\chi$ of indices.  We choose an optimal truncation for the summation over the  index pair $\{\alpha,\beta\}$, which  is given by the singular value decomposition of $M_{A\alpha\beta}$:
\ba\label{svd}
M_{A\alpha\beta} = \sum_{i=1}^{\chi^2} U_{Ai} \lambda_i V_{i \, \alpha\beta}
\ea 
where $\lambda_1\geq \lambda_2\geq\ldots\geq \lambda_{\chi^2}\geq 0$ are positive, and $U,V$ are unitary matrices. The truncation consists then in dropping the smaller set of singular values $\lambda_i$ with $i > \chi$.   Pictorially $V_{i\alpha\beta}$ restricted to $i\leq \chi$ defines a three--valent vertex and we can use these three--valent vertices as in figure \ref{fig:matrix-svd} to arrive at a coarse grained region with less  boundary data.

\subsection{Embedding maps and truncations}

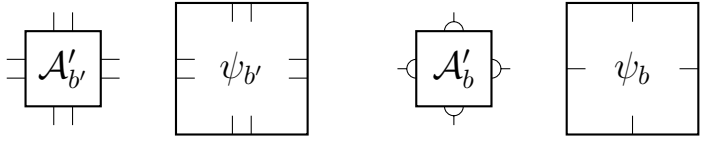
\begin{figure}[h]
\begin{center}
\begin{tikzpicture}[scale=0.5]
\draw [thick] (0,0) rectangle (2,2)
              (4,-0.75) rectangle (7.5,2.75);
\draw (0.75,0) -- (0.75,-0.5)
      (1.25,0) -- (1.25,-0.5)
      (2,0.75) -- (2.5,0.75)
      (2,1.25) -- (2.5,1.25)
      (0,0.75) -- (-0.5,0.75)
      (0,1.25) -- (-0.5,1.25)
      (0.75,2) -- (0.75,2.5)
      (1.25,2) -- (1.25,2.5)
      (1,1) node {\Large $\mathcal{A}'_{b'}$}
      (4,0.75) -- (4.5,0.75)
      (4,1.25) -- (4.5,1.25)
      (7.5,0.75) -- (7,0.75)
      (7.5,1.25) -- (7,1.25)
      (5.5,-0.75) -- (5.5,-0.25)
      (6,-0.75) -- (6,-0.25)
      (5.5,2.75) -- (5.5,2.25)
      (6,2.75) -- (6,2.25)
      (5.75,1) node {\Large $\psi_{b'}$};
\end{tikzpicture}
\quad \quad \quad
\begin{tikzpicture}[scale=0.5]
\draw [thick] (0,0) rectangle (2,2)
              (4,-0.75) rectangle (7.5,2.75);
\draw (1,-0.25) -- (1,-0.5)
      (2.25,1) -- (2.5,1)
      (-0.25,1) -- (-0.5,1)
      (1,2.25) -- (1,2.5)
      (0.75,0) arc(-180:0:0.25)
      (2,0.75) arc(-90:90:0.25)
      (0,0.75) arc(-90:-270:0.25)
      (0.75,2) arc(180:0:0.25)
      (1,1) node {\Large $\mathcal{A}'_{b}$}
      (4,1) -- (4.5,1)
      (7.5,1) -- (7,1)
      (5.75,-0.75) -- (5.75,-0.25)
      (5.75,2.75) -- (5.75,2.25)
      (5.75,1) node {\Large $\psi_{b}$};
\end{tikzpicture}
\caption{To obtain an effective amplitude map $\mathcal{A}'_{b}$ on the coarser boundary $b$ one can pull back the amplitude $\mathcal{A}'_{b'}$ from the finer boundary $b'$ via the previously defined embedding maps.  \label{fig:ampl}}
\end{center}
\end{figure}

We can understand the tensors $V$ as coarse graining maps. Alternatively, if read in the other direction, these maps provide the embeddings $\iota_{bb'}$ from a coarser to a finer discretization.  This interpretation comes from seeing the partition function (with boundary) as a functional (or amplitude map) ${\cal A}_b$ on a `boundary' Hilbert space ${\cal H}_b$. Gluing several space time regions together we obtain a partition  functional ${\cal A}'_{b'}$ which a priori acts on a Hilbert space ${\cal H}_{b'}$ with finer boundary, see figure \ref{fig:ampl}. We can however pull back this functional with the embedding map defined via the tensors $V$ and in this way obtain an effective amplitude map ${\cal A}'_{b}$:
\ba\label{FP}
{\cal A'}_b(\psi_b)&:= & {\cal A'}_{b'} (\iota_{bb'}(\psi_b)) \q .
\ea
This gives a renormalization flow for the amplitude maps, and a fixed point is reached if ${\cal A'}_b={\cal A}_b$.

We see that it is essential to construct three--valent vertices (with its associated tensors), which we can see as special cases of coarse graining or refining maps. These three--valent vertices should be adapted to the four--valent ones giving the `regular' dynamics. They can be understood to give a coarse graining or refining version of the regular evolution defined by the four--valent vertices. Note that this adaptation has to happen after each of the coarse graining steps, as the four--valent tensors flow under coarse graining. 

The singular value decomposition (or generalizations as in \cite{others}) provides one method to construct such three--valent tensors. Interestingly, geometric theories such as, spin foams \cite{alexreview} or spin nets \cite{sffinite,eckert} provide already descriptions for vertices of arbitrary valence \cite{reisenberger,warsaw}.  Thus one can imagine a lattice of say four--valent and three--valent vertices, which automatically implements the coarse graining procedure. However, if one believes that these vertices provide good truncations, in the sense of approximating the summation over an index pair well by a summation over just one index, one needs to adapt the three--valent vertices to the dynamics encoded in the four--valent vertices. That is the embedding maps have to flow together with the effective amplitude maps. Such a relation is provided by the singular value decomposition in (\ref{svd}). Alternatively, spin foam construction tools \cite{reisenberger,warsaw,bdetal13b} provide methods of how to build vertices of arbitrary valency out of a given vertex. This construction has also to be performed at every step of the coarse graining. It will be interesting to see, whether such a method gives a good truncation. The advantage of such a method is that it might be far easier to implement for spin foams, than the singular value decomposition, and might lead to a closed flow equation.  %Tensor network description for spin foams are available, see \cite{eckert}, but involve two types of different tensors. There is also a flow equation available for spin nets \cite{bdetal13a,bdetal13b}, which are analogue models to spin foams. It however involves the singular value decomposition, and hence makes analytical investigations hard. Here a replacement of the singular value construction by some other embedding maps can lead to a flow equation, which is more tractable analytically. 

%Related to this it would be also worthwhile to analyze classically, what kind of projectors are imposed by the (classical) time evolution maps, discussed in section \ref{class}. As discussed there the degrees of freedom that one expects to be projected out quantum mechanically are encoded in the pre--constraints for the coarse graining time evolution map or in the post--constraints for the refining evolution map.

\subsection{Embedding maps for the fixed points}

Ultimately one would expect that the relation between vertices of different valencies is just given by gluing, i.e. a four--valent vertex is given as a gluing of two three--valent vertices. Indeed this relation can be obtained from the singular value decomposition (\ref{svd}) in case that (a) all non--vanishing singular values are equal to one and (b) if one is working in a truncation, the number of non--vanishing singular values needs to be smaller than $\chi$. These conditions are satisfied at the stable fixed points of the renormalization flow (describing the phases of a given system), see for instance \cite{bdetal13a}. 

Condition (a) is expected to arise in theories with diffeomorphism symmetry -- where time evolution is a projector. (Again the problem is that diffeomorphism symmetry is broken under discretization, so the projector property does not hold exactly and might be rather expected to emerge after sufficient coarse graining. For a computation of the transfer matrix in spin foam theories and a discussion whether these are projectors, see \cite{sffinite,bahretal12}.)  

Condition (b), in case that one is working with a cut--off, basically imposes a topological theory for  fixed points that are triangulation invariant\footnote{Rank three tensors can be found from a (singular value) decomposition of the rank four ones. These tensors are associated to three--valent vertices dual to triangles, we can therefore consider models on irregular triangulations. The fixed point condition for coarse graining on a regular lattice are then weaker than requiring triangulation invariance.}.  For instance fixed points identified in \cite{eckert,bdetal13a,bdetal13b} via a tensor network coarse graining describe triangulation invariant and therefore topological models and $\chi$ gives the maximal number of propagating degrees of freedom. Indeed we will see in section \ref{top} that all the proposals outlined here are explicitly realized. 
For an interacting theory, such as 4D gravity, one would expect to need an infinite bond dimension $\chi$, as indeed arises around phase transitions. With a fixed $\chi$ one can however approach the phase transition up to a certain precision, and, as the method is designed to keep  the variables describing coarse excitations,  one can hope to obtain reliable predictions for sufficiently coarse observables. In light of the previous discussion this means to obtain amplitude functionals, which satisfy the cylindrical consistency conditions sufficiently well for coarse boundaries.
%To deal with such a situation, we believe that one needs to employ the formalism developed in section \ref{howto}, which allows infinite bond dimension, i.e. infinitely refined boundaries.

\section{Topological theories}\label{top}

The previously introduced and discussed concepts of time evolution via coarse graining / refining and the concept of  cylindrically consistent amplitudes are perfectly realized in topological field theories. In the following we would like to emphasize a few key points. In the first part of this section, we will mainly refer to topological lattice field theories in 2D, for instance \cite{fuku}. For theories with a geometric interpretation, see for instance \cite{bdwk, bdetal13b}.

Consider a 2D lattice topological field theory with partition functions defined on three--valent graphs. As for a  three--valent tensor network we have weights or tensors associated to the vertices and variables and hence Hilbert spaces associated to the edges. The partition function is then defined by summing the variables associated to bulk edges. In case of a boundary, we keep the corresponding variables fixed, thus obtaining a partition function depending on these boundary values. Alternatively we can understand the partition function as an operator (between two boundary Hilbert spaces) or a function (on one boundary Hilbert space). 

In this vain a three--valent graph 
\ba
\begin{tikzpicture}[baseline,scale=0.75]
\draw (-0.5,-0.5) -- (0,0) -- (0.,0.5)
      (0.5,-0.5) -- (0,0.0);
\end{tikzpicture}
\ea
represents the simplest graph and the associated partition function, interpolating between a two--site boundary Hilbert space on the lower boundary and a one--site boundary Hilbert space on the upper boundary. In this case we can understand this partition function as a (coarse graining) time evolution map, not involving any bulk summation. The same holds for the time inverted vertex.

Topological lattice theories are triangulation invariant (referring to the triangulation dual to the graph). Thus the partition function does only depend on the topology of the manifold and not on the choice of the triangulation. (We will later show that this also holds in a certain sense for the boundary triangulation, due to the cylindrical consistency of the partition function.) Pictorially this corresponds to the following equalities
\ba
\begin{tikzpicture}[baseline,scale=0.7]
\draw (-1,-0.75) -- (-0.5,-0.25) -- (0,0.25) -- (0,0.75)
      (-0.5,-0.25) -- (0.5,-0.25) -- (0,0.25)
      (0.5,-0.25) -- (1,-0.75);
\end{tikzpicture}
\; = \; c
\begin{tikzpicture}[baseline,scale=0.7]
\draw (-1,-0.75) -- (0,0) -- (0,0.75)
      (1,-0.75) -- (0,0);
\end{tikzpicture} \quad , \quad
\begin{tikzpicture}[baseline,scale=0.7]
\draw (-1,0.75) -- (-0.5,0.25) -- (0,-0.25) -- (0,-0.75)
      (-0.5,0.25) -- (0.5,0.25) -- (0,-0.25)
      (0.5,0.25) -- (1,0.75);
\end{tikzpicture} \; = \;
c
\begin{tikzpicture}[baseline,scale=0.7]
\draw (-1,0.75) -- (0,0) -- (0,-0.75)
      (1,0.75) -- (0,0);
\end{tikzpicture}
\quad , \quad
\begin{tikzpicture}[baseline]
\draw (-0.5,-0.5) -- (-0.5,0.5)
      (-0.5,0) -- (0.5,0) -- (0.5,0.5)
      (0.5,-0.5) -- (0.5,0);
\end{tikzpicture}
\; = \;
\begin{tikzpicture}[baseline]
\draw (-0.5,-0.5) -- (0,-0.25) -- (0,0.25) -- (-0.5,0.5)
      (0.5,-0.5) -- (0,-0.25)
      (0.5,0.5) -- (0,0.25);
\end{tikzpicture} \quad .
\ea
Here we assume always a summation over the variables or indices associated to the bulk edges. The equations have to hold for all possible choices of the boundary variables.
In this section we will assume that the constant $c$ is actually finite and hence can be adjusted to $c=1$ by a rescaling of the amplitudes. 
%Note however that in gravitational theories with non--compact diffeomorphism symmetry, this constant is related to the size of the gauge orbit, and hence equals infinity. 

Given the $2-2$ move invariance, we can replace the $3-1$ move by the so--called bubble move 
\ba\label{bubble}
\begin{tikzpicture}[baseline,scale=0.75]
\draw (0,-1) -- (0,-0.5) -- (-0.5,0) -- (0,0.5) -- (0,1)
      (0,-0.5) -- (0.5,0) -- (0,0.5)
      (0,-1.3) node {$j$}
      (0,1.3) node {$j$};
\end{tikzpicture} \; = \;
c \; \;
\begin{tikzpicture}[baseline,scale=0.75]
\draw (0,-1) -- (0,1)
      (0.25,0) node {$j$};
\end{tikzpicture} \quad .
\ea

The equivalence of bubble and $3-1$ move (given the $2-2$ move holds) follows from the following calculation
\ba
\begin{tikzpicture}[baseline,scale=0.7]
\draw (-1,-0.75) -- (-0.5,-0.25) -- (0,0.25) -- (0,0.75)
      (-0.5,-0.25) -- (0.5,-0.25) -- (0,0.25)
      (0.5,-0.25) -- (1,-0.75);
\end{tikzpicture} \; = \;
\begin{tikzpicture}[baseline,scale=0.75]
\draw (-1,-0.75) -- (-0.5,-0.5) -- (-0.5,0.) -- (-1,0.25) -- (-0.5,0.5) -- (-0.5,0.75)
      (-0.5,-0.5) -- (0,-0.75)
      (-0.5,0) -- (0,0.25) -- (-0.5,0.5);
\end{tikzpicture}\; = \;
c
\begin{tikzpicture}[baseline,scale=0.7]
\draw (-1,-0.75) -- (0,0) -- (0,0.75)
      (1,-0.75) -- (0,0);
\end{tikzpicture}
\quad .
\ea

\subsection{Time evolution as coarse graining and refining}

Using this pictorial representation we can symbolize a local time evolution operator or transfer matrix acting on a two  site boundary Hilbert space  as follows:
\begin{equation}
T \; = \;
\begin{tikzpicture}[baseline,scale=0.75]
\draw (-0.5,-1) -- (0,-0.5) -- (0,0.5) -- (-0.5,1)
      (0.5,-1) -- (0,-0.5)
      (0.5,1) -- (0,0.5);
\end{tikzpicture}
\; = \; \sum_i \lambda_i  \left| \iota^{(i)} \right\rangle \left \langle \iota^{(i)} \right| \quad .
\end{equation}
Time is flowing upward. From the bubble move we see that the time evolution operator  is actually a projector: 
\begin{equation}
T^2 \; = \;
\begin{tikzpicture}[baseline,scale=0.75]
\draw (-0.5,-1) -- (0,-0.75) -- (0,-0.25) -- (-0.5,0) -- (0,0.25) -- (0,0.75) -- (-0.5,1)
      (0.5,-1) -- (0,-0.75)
      (0,-0.25) -- (0.5,0) -- (0,0.25)
      (0,0.75) -- (0.5,1);
\end{tikzpicture}
\; = \; \sum_{i,j} \lambda_i \lambda_j \; \underbrace{\left\langle \iota^{(i)} \right. \left | \iota^{(j)} \right \rangle}_{=\delta_{i,j}} \; \left |\iota^{(i)} \right \rangle \left \langle \iota^{(j)} \right | \; = \; \sum_i \lambda_i^2 \left |\iota^{(i)} \right \rangle \left \langle \iota^{(i)} \right | \; = \;
\begin{tikzpicture}[baseline,scale=0.75]
\draw (-0.5,-1) -- (0,-0.5) -- (0,0.5) -- (-0.5,1)
      (0.5,-1) -- (0,-0.5)
      (0.5,1) -- (0,0.5);
\end{tikzpicture}
\; = \;
T \quad .
\end{equation}
Hence the eigenvalues are  $\lambda_i=1 \vee \lambda_i=0$. The construction of dynamical embedding maps via a singular value decomposition is trivial and it is straightforward to split the projector into two maps, one that can be interpreted as a coarse graining, the other in terms of a refining:
\begin{equation}
C\; := \;
\begin{tikzpicture}[baseline,scale=0.75]
\draw (-0.5,-0.5) -- (0,0) -- (0.,0.5)
      (0.5,-0.5) -- (0,0.0);
\end{tikzpicture}
\quad , \quad 
R \; := \;
\begin{tikzpicture}[baseline,scale=0.75]
\draw (0,-0.5) -- (0,0) -- (-0.5,0.5)
      (0,0) -- (0.5,0.5);
\end{tikzpicture} \quad .
\end{equation}
Each of these maps can be interpreted as maps between Hilbert spaces of different dimension. Concatenating the two  gives either the time evolution operator back or is just the identity: This tells us both that the refined state does not carry additional information and that no (physical) information is lost under coarse graining. We will extend this to arbitrarily large triangulations below.

The time evolution operator acts locally, such that it is possible to only locally evolve a state in time, e.g.:
\begin{equation}
\begin{tikzpicture}[baseline,scale=0.75]
\draw (-0.5,-1) -- (0,-0.5) -- (0,0.5) -- (-0.5,1)
      (0.5,-1) -- (0,-0.5)
      (0.5,1) -- (0,0.5)
      (1,-1) -- (1,1)
      (1.5,-1) -- (1.5,1);
\end{tikzpicture} \quad .
\end{equation}
However in which order one time evolves pairs of lattice sites is an arbitrary choice, one which should not influence the results of the theory such as the partition function. Therefore we impose the following consistency condition, which is satisfied in topological field theories and allows us to define the transfer matrix for three discretization sites uniquely:
\begin{equation}\label{6.9}
\begin{tikzpicture}[baseline,scale=0.6]
\draw (-0.5,-2) -- (0,-1.5) -- (0,-0.5) -- (-0.5,0) -- (-0.5,2)
      (0.5,-2) -- (0,-1.5)
      (0.5,0) -- (0,-0.5)
      (1.5,-2) -- (1.5,0)
      (0.5,0.) -- (1,0.5) -- (1,1.5) -- (0.5,2)
      (1.5,0.) -- (1,0.5)
      (1.,1.5) -- (1.5,2);
\end{tikzpicture}
\; = \;
\begin{tikzpicture}[baseline,scale=0.6]
\draw (0.5,-2) -- (1,-1.5) -- (1,-0.5) -- (0.5,0) 
      (-0.5,-2) -- (-0.5,0)
      (1.5,-2) -- (1,-1.5)
      (1.5,0) -- (1,-0.5)
      (1.5,0) -- (1.5,2)
      (-0.5,0.) -- (0,0.5) -- (0,1.5) -- (-0.5,2)
      (0.5,0.) -- (0,0.5)
      (0.,1.5) -- (0.5,2);
\end{tikzpicture}
 \; = \;
\begin{tikzpicture}[baseline,scale=0.6]
\draw (-0.5,-2) -- (0,-1.5) -- (0.5,-2)
      (0,-1.5) -- (0.5,-1) -- (1.5,-2)
      (0.5,-1) -- (1.5,0) -- (0.5,1) -- (1.5,2)
      (0.5,1) -- (0,1.5) -- (0.5,2)
      (0,1.5) -- (-0.5,2);
\end{tikzpicture} \quad .
\end{equation}
This construction can be generalised to  arbitrarily many discretization sites. In all cases we can replace the time evolution map with a graph of the form on the right hand side of equation (\ref{6.9}).  Thus the maximal rank of this time evolution map (which is a projector) is given by the bond dimension of the kink in the middle, i.e. the bond dimension of one edge. This gives also the maximal number of physical degrees of freedom.

\subsection{Consistent embedding maps and inductive limit}

Furthermore if one cuts this diagram into two pieces at the kink, one obtains both more general coarse graining and refining maps.  Hence the consistency conditions naturally translate to both the coarse graining and refining maps, as we demonstrate for the refinement map $R$:
\begin{equation} \label{eq:mps}
\begin{tikzpicture}[baseline,scale=0.75]
\draw (-0.5,-1) -- (-0.5,-0.5) -- (0,0) -- (0,0.5) -- (0.5,1)
      (-0.5,-0.5) -- (-1,0)
      (0,0.5) -- (-0.5,1);
\end{tikzpicture}
\; = \;
\begin{tikzpicture}[baseline,scale=0.75]
\draw (-0.5,-1) -- (-0.5,-0.5) -- (-1,0) -- (-1,0.5) -- (-1.5,1)
      (-0.5,-0.5) -- (0,0)
      (-1,0.5) -- (-0.5,1);
\end{tikzpicture} \; = \;
\begin{tikzpicture}[baseline,scale=0.75]
\draw (1.,-0.5) -- (0.5,0) -- (1.5,1)
      (0.5,0) -- (0,0.5) -- (0.5,1)
      (0,0.5) -- (-0.5,1);
\end{tikzpicture}
\quad .
\end{equation}

Thus these refinement maps satisfy the path independence conditions as outlined in section \ref{pathdep}, and therefore allow the construction of an inductive  limit of Hilbert spaces as described in section \ref{emb}. This gives the continuum limit of this theory. Furthermore, understanding the partition function ${\cal A}_b$ (with boundary $b$) as a functional on the boundary Hilbert space
\ba
{\cal A}_b: {\cal H}_b \rightarrow \mathbb{C}\q ,
\ea
we also obtain that the partition function is a cylindrically consistent operator \cite{bdwk}: 
\ba
{\cal A}_{b'}(\iota_{bb'}(\psi_b)) &=& {\cal A}_b(\psi_b)
\ea
(which coincides with the fixed point condition \eqref{FP}). Here $\iota_{bb'}$ is built from the refinement maps $R$, in the way described above.
Given a boundary $b$ we choose a triangulation  (or dual three--valent graph)  interpolating this boundary. 
As long as we choose a fixed topology for this interpolating triangulation, the partition function will not depend on this choice and hence is a well defined functional on the boundary Hilbert space. Additionally we can refine the boundary via a refinement move. An interpolating triangulation can be obtained by just including a coarse graining move at the appropriate dual edge. This will give a `bubble' that can be removed due to the bubble move invariance, and we arrive at the previous partition function acting on the unrefined Hilbert space. 

Hence the partition function (actually a family of functionals labelled by the boundaries $b$) is cylindrically consistent with respect to the embeddings provided by the refining time evolution. That automatically allows to define from the (so far) discrete partition functions a continuum limit on the projective limit Hilbert space.  This is to our knowledge a new insight, as topological theories are often only discussed with regard to the invariance of the bulk triangulation.

The refinement maps can also be used to construct the Hartle Hawking vacuum states mentioned in section \ref{class}. To this end one has either to dualize one edge (graphically a bent or cup). Alternatively in examples where the edges are labelled by $(SU(2))$ spins, we start with a refining map for which we fix on the incoming edge $j=0$, which gives the Hilbert space $\mathbb{C}$ associated to this edge. The (two--site) Hilbert space can then be refined further in an arbitrary way, giving the Hartle Hawking vacuum state on boundaries with different numbers of sites.

Doing this in a linear way, i.e. as for the graph on the right in equation (\ref{eq:mps}) 
gives   matrix product states (MPS) \cite{mps}. MPS provide ans\"atze for ground state wave functions of Hamiltonians. The projectors $T$ defined above  correspond to exponentiated Hamiltonians and the type of MPS defined in \eqref{eq:mps} is the ground state to the following Hamiltonian:
\begin{equation}\label{ham}
H = \sum_{I=1}^{N-1} \left(\mathbb{I} - {
\begin{tikzpicture}[baseline,scale=0.5]
\draw (-0.5,-1) -- (0,-0.5) -- (0,0.5) -- (-0.5,1)
      (0.5,-1) -- (0,-0.5)
      (0.5,1) -- (0,0.5);
\end{tikzpicture}}_I \right) \quad ,
\end{equation}
where $I$ denotes the index pair the projector is acting on for a total of $N$ outgoing legs. Thus Hartle Hawking vacuum states appear here as the ground states of the Hamiltonians (\ref{ham}), justifying again the notion of these states as vacuum states.

\subsection{3D topological theories and entangling moves}

Similar statements hold for higher dimensional theories, for instance $BF$ theories. There are however interesting differences pertaining to the role of Pachner moves in discrete topological theories based on triangulations, such as the Turaev--Viro models \cite{tuarev}. The physical states of these models can be described as string net states \cite{stringnets}.

For a canonical time evolution in (2+1)D we will have  $3-1$, $1-3$ and $2-2$  moves as time evolution moves as described in section \ref{class}. 
These allow 
 to build an arbitrary complex triangulated hypersurface from a simple one.  In this way one can build up an analogous MPS representation of a string net state on arbitrary complex 2D triangulations or on the corresponding dual graphs.  The $3-1$ and $1-3$ moves serve as (purely) coarse graining or refining moves, whereas the $2-2$ moves  (dis--)entangle the degrees of freedom. The latter play an important role in entanglement renormalization \cite{vidal,koenig}.

Interestingly MPS states in one (spatial) dimensions do not lead to long range entanglement \cite{schuch}, whereas the example just described gives a phase with long range entanglement in two (spatial) dimensions \cite{koenig}. This might be due to the necessity of the $2-2$ move to obtain triangulations not equivalent to a stacked sphere (which would not support long range entanglement). In the case of the stacked sphere the consecutive $1-3$ moves can be represented by a  tree graph.  In (1+1)D all triangulations of a circle can be obtained  as `stacked spheres', which are dual to trees.

In $(3+1)$D we have similarly $1-4$  and  $4-1$ moves as refining and coarse graining moves respectively. As mentioned before the $4-1$ move does not add physical degrees of freedom, as all additional degrees of freedom are associated to Hamiltonian and diffeomorphism constraints \cite{bahrdittrich09a,hoehn2}. Additionally we have  $2-3$ and $3-2$ moves, which can  be interpreted as entangling moves, similarly to the $2-2$ move in $(2+1)$D.

\subsection{Constructing inductive limit Hilbert spaces for non--topological theories}\label{subtop}

We discussed that the embedding maps provided by the refining time evolution of topological theories satisfy the consistency conditions (\ref{cons1}). Thus one can use these embeddings to construct an inductive limit Hilbert space as outlined in section \ref{emb}.

Note that this Hilbert space will support a much bigger class of observables than just the Dirac, i.e. topological observables of the topological theory. The set of observables supported by this Hilbert space is determined by the cylindrical consistency conditions (\ref{obs}) for observables. The cylindrical consistent observables then describe excitations from the vacuum state, which is given by the no--boundary wave function of the topological theory.
Thus the excitations in particular violate the constraints (implementing the equations of motion) of the topological theory. 

This is a general proposal for the construction of inductive Hilbert spaces. It will be interesting to explore more in detail the relation between the set of cylindrical observables which characterize the inductive Hilbert space and the topological theory which provides the embedding maps. 

This strategy to construct an inductive limit Hilbert space has been realized recently \cite{geiller} for the topological $BF$ theory, which describes the moduli space of flat connections. In this case the excitations are parametrized by curvature observables. The set of cylindrically consistent observables is given by a holonomy--flux algebra underlying the formulation of loop quantum gravity and (lattice) gauge theories. This method therefore resulted in an alternative representation and a new vacuum for loop quantum gravity.

\section{Geometric interpretation of the refining maps}\label{geom}

Here we wish to point out that geometric theories are very special with regards to coarse graining and refining.\footnote{See also the discussion in \cite{ditt}, which argues that in reparametrization invariant theories all couplings are dimensionless.} This is due to the fact that the geometry itself is included into the set of dynamical variables. The (say semi--classical) state on the boundary of a region determines a geometry for the bulk, defined as solution of the Einstein equations for the given boundary data the state is peaked on. (This of course assumes that one has a sensible theory of quantum gravity, which would result in a semi--classical state for the Hartle Hawking state.) Thus setting (geometric) scale and number of coarse graining steps as equal is, at least a priori, senseless in such theories. Rather, a renormalization scale is given by the coarseness or fineness of the boundary data, that is, the scale on which geometric properties, such as curvature, vary. 

Even if one peaks the boundary state on a given geometry with a fixed (hypersurface) volume one cannot expect to find that the partition functions peaks on some regular bulk geometry such that the bulk volumes are bounded by the hypersurface volume.

The reason is that one expects diffeomorphism symmetry to emerge in the form of vertex translation invariance. This symmetry even allows to move the vertices such that orientations of building blocks are inverted. This corresponds to `spikes' in the geometry, see for instance \cite{orient,riello}. These spikes give rise to divergences \cite{louapre,smerlak,riello,bonzom}, related to the non--compactness of the diffeomorphism gauge orbits.  As we will argue below this mechanism allows the appearance of arbitrarily large spins even in a region bounded by a small boundary geometry. This might make even a theory describing flat geometry, such as 3D $BF$, appear as highly fluctuating. However (almost) all these fluctuations are gauge fluctuations \cite{smerlak}, due to the diffeomorphism gauge symmetry. We will illustrate this with a 2D example below.

The relation between the sum of orientations and divergences has been pointed out in \cite{orient} which also argues that allowing only one orientation could cure the problem of divergences. However, we will show here, that from the perspective of time evolution as a refining and coarse graining map, the appearance of two orientations is very natural. (It is also natural as the gravitational constraints are quadratic in the momenta describing time symmetric evolution. The two solutions of the quadratic equation correspond to the two orientations.)

%That the time evolution for spin foams, in the form of gluing simplices to a hypersurface, involves both orientations (forward and backward in time) follows also from a semi--classical analysis of the simplex amplitude.
 %In such an expansion, which are performed via a stationary phase approximation in a large spin limit \cite{asympt}, one obtains the Regge action associated to the respective simplex as the dominating phase of the path integral. Yet instead of $e^{i S_R }$ one  ends up with the cosine $\cos(S_R)= \tfrac{1}{2}(e^{i S_R }+e^{-i S_R })$. Such an expansion is also possible for the intertwiner models introduced in \cite{bdwk}, as the fixed point amplitude for these models results in $\{6j\}$ and Clebsch--Gordan symbols.  
 
\subsection{2D example}

Let us illustrate this with the intertwiner models introduced in \cite{bdwk}. These host families of topological theories, which have all the properties discussed in section \ref{top}. Moreover the theories allow for a natural geometric interpretation, as they are defined on three--valent graphs. The edges carry a spin $j$ ($SU(2)$ representation) and a magnetic quantum number. The spin can be interpreted as a length variable -- indeed at the three--valent vertices triangle inequalities have to be satisfied arising from $SU(2)$ recoupling theory.

Consider one upward pointing line as in the examples before: This line can be interpreted as a line in a (2$D$) space with a length given by the spin $j$. Using our previously defined refining maps $R$, we can map it to a different state, which is now labelled by two spins $j'$ and $j''$: 
\begin{eqnarray}
\begin{tikzpicture}[baseline]
\draw[|-|] (-0.5,0.1) -- (0.5,0.1);
\draw      (0,-0.1) node {$j$};
\end{tikzpicture} \quad &\quad \rightarrow \quad & \quad \sum_{j}\;
\begin{tikzpicture}[baseline]
\draw[|-|] (-0.5,0.1) -- (0.5,0.1);
\draw      (0,-0.1) node {$j$};
\draw (-0.5,0.1) -- (0,0.6) -- (0.5,0.1)
      (-0.3,0.5) node {$j'$}
      (0.4,0.5) node{$j''$};
\end{tikzpicture} \\
\begin{tikzpicture}[baseline]
\draw (0,-0.3) -- (0.,0.3);
\end{tikzpicture}\quad &\quad \rightarrow \quad & \quad
\begin{tikzpicture}[baseline]
\draw (0,-0.5) -- (0,0) -- (0.5,0.5)
      (0,0) -- (-0.5,0.5);
\end{tikzpicture}  \quad ,
\end{eqnarray}
where the spins $j'$ and $j''$ have to satisfy triangle inequalities. However this picture is indistinguishable from adding a triangle with opposite orientation and hence `removing' it: 
\begin{equation}
\begin{tikzpicture}[baseline]
\draw[|-|] (-0.5,0.1) -- (0.5,0.1);
\draw      (0,-0.1) node {$j$};
\end{tikzpicture}
\quad \rightarrow \quad 
\begin{tikzpicture}[baseline]
\draw (-0.5,0.1) -- (0,-0.4) -- (0.5,0.1)
      (-0.3,-0.4) node {$j'$}
      (0.4,-0.4) node{$j''$};
\end{tikzpicture} \quad.
\end{equation}
Thus for a quantum theory this means that both possible orientations have to be taken into account in a superposition:
\begin{equation}
\begin{tikzpicture}[baseline]
\draw[|-|] (-0.5,0) -- (0,0);
\draw[|-|] (0,0) -- (0.5,0.);
\end{tikzpicture}
\; \sim \;
\begin{tikzpicture}[baseline]
\draw[|-|] (-0.5,0) -- (0.5,0);
\draw (-0.5,0.) -- (0,0.5) -- (0.5,0.);
\end{tikzpicture}
\; \oplus \;
\begin{tikzpicture}[baseline]
\draw[|-|] (-0.5,0) -- (0.5,0);
\draw (-0.5,0.) -- (0,-0.5) -- (0.5,0.);
\end{tikzpicture} \quad .
\end{equation}

This picture conforms with refining the edge and adding a vacuum degree of freedom (this degree of freedom is not physical, as we are considering a topological theory here): This vacuum degree of freedom allows fluctuations of the edge geometry around a flat subdivision -- in the sense that the refined edge can bend upwards or downwards. Any asymmetry would appear as proper time evolution, which we do not expect for a topological  or gravitational theory. In this case time evolution is generated by constraints and hence gauge -- and as explained before realized as a projector in the quantum theory.

 Note that the fluctuation can be arbitrarily large, as argued below. In this case this can be linked to diffeomorphism symmetry realized by a vertex translation symmetry. The middle vertex can be translated an arbitrary large distance forward or backward (or sideways) in  `time'.  Moreover, as this is a gauge symmetry, all such configurations have to be gauge equivalent, i.e. come with the same amplitude (and a diffeomorphism invariant measure\footnote{This actually allows to determine the path integral measure, see \cite{sebmeas}.}).

\subsection{Spin foams}

A similar picture applies to spin foams, where gluing a simplex to a boundary can be done with two orientations. From the semi--classical expressions for the simplex amplitude we again obtain a geometric picture: i.e. with a $1-4$ move (from gluing a 4--simplex to one boundary tetrahedron) we replace a boundary tetrahedron with a complex of four tetrahedra, that now allows the inner geometry of the original tetrahedra to fluctuate around a flat subdivision. Note that also in this case one did actually not add a physical degree of freedom, at least not if one deals with Regge geometries \cite{ryan,hoehn1,hoehn2}. The reason is that the new kinematical degrees of freedom (the four new edge lengths) are accompanied by four (Hamiltonian and diffeomorphism) constraints, associated to the new vertex. These allow to move the additional vertex arbitrarily forward or backward in time, explaining the appearance of the two orientations. As before the diffeomorphism or vertex translation symmetry also means that configurations with arbitrarily large length of the four inner edges have equal amplitude to those describing a 'flat' subdivision, thus one would expect divergences to appear for every inner vertex, for a discussion in spin foams see \cite{louapre,smerlak,claudio,riello,bonzom}. Thus one should be very careful with treating the spin $j$ variable, which encode the length or area variables in 3D or 4D respectively, as an order parameter. (Indeed one should consider diffeomorphism invariant observables as order parameters, which are however hard to come by \cite{partial}.)

We can provide here an interpretation of the divergences as coming from (extremely) squeezed states: As mentioned we add only degrees of freedom in the vacuum state (including gauge degrees of freedom), in the case of $4-1$ moves  these have to satisfy the Hamiltonian and diffeomorphism constraints. Thus, fluctuations in the `constraint' directions are completely suppressed, whereas fluctuations in the conjugated (i.e.\ gauge) directions become infinitely large, represented as a non--compact gauge orbit of configurations with equal weight.

\section{Discussion}\label{discuss}

We pointed out that gravitational theories in a  simplicial description, provide with their time evolution maps automatically refining, coarse graining and entangling maps. %This applies also to more general discretizations, i.e. involving other building blocks. This notion generalizes also to theories defined on a fixed space time background, i.e. the massless scalar field. 
%
%The two cases of geometric and non--geometric theories differ however in the following: In the first case time evolution is (approximately) implemented  as a projector whereas in the second case it includes a proper notion of (forward) time evolution. Applying a refining time evolution move  in a gravitational theory we can interpret the state as a superposition of a forward and backward evolved one, see the example in section \ref{subscalar}.  In the quantum theory that leads to a `squeezed' state, in the sense that constraints, originating from diffeomorphism symmetry, have to be satisfied exactly. Thus the added degrees of freedom fluctuate widely in the conjugate, i.e. gauge direction.
%
%
More generally we interpret the degrees of freedom added during refining time evolution moves as degrees of freedom in the vacuum state (or gauge degrees of freedom). 
This suggests the construction of a global vacuum state as a state evolved from a one--dimensional Hilbert space ${\mathbb C}$, see also \cite{hoehn2,hoehnq}, which gives a simplicial realization of   the no--boundary proposal \cite{hhv}. Indeed via the notion of dynamical cylindrical consistency \cite{bd12b}, we can identify the vacuum states as representing the equivalence class which includes the unique state in the `no--boundary' Hilbert space ${\mathbb C}$  on different discretizations.

We argued that the time evolution maps  provide embedding maps for the construction of the continuum limit via projective techniques. %The latter is standard in loop quantum gravity, however so far based on the kinematical Ashtekar--Lewandowski vacuum and kinematical embedding maps. 
In section \ref{howto} we outlined  how to define and arrive at a consistent continuum dynamics for quantum gravity. This is based on the dynamical embedding maps and proposes to construct the amplitude maps as cylindrically consistent maps based on these embeddings. This allows to define the amplitude maps as objects of the continuum theory on the continuum Hilbert space.

Such (dynamical) embedding maps have however to satisfy stringent path independence conditions, which we related to the path independence under different choices of interpolating hypersurfaces \cite{kuchar} and an anomaly free representation of the Dirac algebra of constraints \cite{thiemann, bonzomDirac}. These conditions are indeed hard to satisfy exactly for interacting theories but should be valid in some approximate sense if considering sufficiently coarse grained observables. 

We explained that tensor network renormalization algorithms provide a method to construct dynamical embedding maps that do satisfy the consistency conditions to a better approximation and the related (approximately) cylindrically consistent amplitudes.
  An important ingredient in these  algorithms are  truncations. Good truncations are basically good reorganizations of the degrees of freedom into coarser ones and finer ones. We argued that such a splitting can be found by employing radial, that is refining, evolution.

In topological theories the refining time evolution maps typically satisfy the path independence conditions. This allows the construction of projective limit Hilbert spaces using refining time evolution as embedding maps. This will realize the physical state of the topological theory (satisfying the constraints of the topological theory) as a vacuum in this projective limit Hilbert space. This vacuum coincides with the no--boundary wave functions. Excitations can be produced by cylindrically consistent observables. An example of this construction has been recently provided in \cite{geiller}.

For non--topological theories, such as 4D gravity, we suggest that an exact satisfaction of the path independence conditions for the embedding maps would rather involve non--local dynamics, as is indicated by the discussion in section \ref{scalar}. The construction of the continuum limit in section \ref{howto} allows for such non--local embeddings. The necessity of a non--local dynamics has been recently argued for in \cite{nonlocal}, which points out that linearized 4D quantum Regge calculus requires a non--local path integral measure in order to show invariance under $5-1$ moves. 

From a statistical physics point of view one would expect that a second order phase transition is needed for the continuum limit, leading to long range (in terms of number of lattice cites) correlations and a conformal theory at the boundary.
Indeed in the context of tensor network algorithms and radial evolution the appearance of a conformal theory at the fixed point leads to an interpretation in terms of  AdS geometries and holographic renormalization, for instance \cite{swingle}. For the case of non--perturbative gravity such an interpretation might not apply straightforwardly. Here one would expect that the boundary variables or the quantum state defined on the boundary encodes the geometry of the boundary and  -- via the equations of motion -- of the bulk. %The  question is, if the Hartle Hawking vacuum state   is peaked on AdS geometry. 

%A naive geometric interpretation of the variables appearing in spin foams or spin nets, might give the impression of highly fluctuating geometries as discussed in section \ref{geom}. We emphasized the interpretation of such fluctuations as gauge degrees of freedom and the importance of specifying gauge invariant order parameters. Thus one can for instance not expect that a refinement of some region will lead to small expectation values for the spins, rather (arbitrary) large spikes or bubbles may form and indeed should receive equal weights to a (roughly) flat subdivision of the region.

%It seems that in this context the entangling moves, i.e. $3-2$ and $2-3$ moves  or $2-2$ moves in $(2+1)$D play an important role of distributing the added vacuum degrees of freedom in a more regular fashion to the entire triangulation. In the context of topological phases, of which $3$D gravity is one example, such moves lead to the generation of long range entanglement \cite{koenig}. 

 There are still many puzzling features to explore in the context of discretization changing time evolution. This in particular applies to interacting theories, such as 4D gravity. As we outlined here such discretization changing evolution might however provide a definition of the physical vacuum and more generally allow the construction of the continuum limit of the theory. This makes the explorations of these issues very worthwhile.

\section*{Acknowledgements}
B.D. thanks Philipp H\"ohn for collaboration, intensive discussions and for providing a draft version of \cite{hoehnq}, as well as Kyle Tate for collaboration on \cite{unpublished}. We thank Wojciech Kaminski and Mercedes Martin-Benito for collaborations and discussions and Guifre Vidal for pointing out reference \cite{koenig} and explaining the (corner transfer) tensor network algorithm to us. B.D. would like to thank Laurent Freidel, Marc Geiller, Ted Jacobson, Aldo Riello and Lee Smolin  for discussions.  We thank the referee for helpful comments, which led to an improvement of the manuscript. S.St. gratefully acknowledges support by the DAAD (German Academic Exchange Service) and would like to thank Perimeter Institute for an Isaac Newton Chair Graduate Research Scholarship. 
This research was supported by Perimeter Institute for Theoretical Physics.
Research at Perimeter Institute is supported by the Government of Canada through Industry Canada and by the
Province of Ontario through the Ministry of Research and Innovation.

%--------------------------------------------------------------------------------------------------

\end{document}